\documentclass{report}
\usepackage{amsmath}
\usepackage{graphicx}
\usepackage{rotating}
\usepackage{epsfig}
\begin{document}
\ifx\href\undefined\else\hypersetup{linktocpage=true}\fi

\def\sinsqthot{$\sin^2(2\theta_{13})$}
\def\dmsqotw{$\Delta m_{12}^2$}
\def\sol{$\Delta m_{12}^2$}
\def\dmsqoth{$\Delta m_{13}^2$}
\def\dm12{\Delta \mbox{m_{12}}^2}
\def\dm13{\Delta m_{13}^2}
\def\nubar{\bar \nu}
\def\dmsq23{\Delta m^2_{32}}
\def\delmsq{\Delta m^2}
\def\numu{\nu_\mu}
\def\nutau{\nu_\tau}
\def\nue{\nu_e}
\def\sinsq2t23{\sin^2 2\theta_{23}}
\def\ssq2t13{\sin^2 2\theta_{13}}
\def\ss2t12{\sin^2 2\theta_{12}}
\def\th13{\theta_{13}}
\def\numutonue{\nu _{\mu }\rightarrow \nu _{e}}
\def\numutotau{\nu _{\mu }\rightarrow \nu _{\tau}}
\def\pizero{\pi^0}
\def\numubar{\bar{\nu}_\mu}
\def\nubar{\bar{\nu}}
\def\nuebar{\bar{\nu}_e}
\def\etal{{\it et al.}}
\begin{titlepage}
\begin{center}
\vspace*{1.0cm}
\vskip 2cm
\vskip 2cm
{\LARGE \bf Letter of Intent\\
\vspace*{6mm}
 FLARE\\
\vspace*{6mm}
 Fermilab Liquid ARgon Experiments }\\
\vskip 1cm
{\it \large Version 1.0}\\     
\vspace*{5mm}
{\it \large \today}\\
\vspace*{1.0cm}
\mbox{Bartoszek~Eng.}~-~Duke~-~Indiana~-~Fermilab~-~LSU~-~MSU~-
~Osaka~-~Pisa~-~Pittsburgh~-~Princeton~-~Silesia~
~\mbox{South Carolina}~-~\mbox{Texas A\&M}~-
~Tufts~-~UCLA~-~\mbox{~Warsaw University}~-
~\mbox{~INS Warsaw}~-~Washington~-~\mbox{York-Toronto}

\end{center}
\end{titlepage}

\newpage
\begin{center}

L.~Bartoszek\\
{\it Bartoszek Engineering}\\
\vspace{5mm}
K. Scholberg\\
{\it Duke University, Durham, NC}\\
\vspace{5mm}

R.~Tayloe,~G.~Visser\\
{\it Indiana University, Bloomington, IN}\\
\vspace{5mm}

B.T.~Fleming,~G.F.~Foster,R.~Hatcher,~W.~Jaskierny,
~H.~Jostlein,~C.~Kendziora,~J.~Kilmer,~K.~Krempetz,
~V.~Makeev,~A.~Marchionni,
~A.~Para$^{\bf{\dag}}$,~S.~Pordes,~P.~Rapidis,
~R.~Schmitt,~Z.~Tang,~B.~Wands \\
{\it Fermilab, Batavia, IL}\\
\vspace{5mm}

W.~Metcalf,~M.0.~Wascko \\
{\it Louisiana State University, Baton Rouge, LA}\\
\vspace{5mm}

C. Bromberg, R. Richards\\
{\it Michigan State University, East Lansing, MI}\\
\vspace{5mm}

H.~Ejiri,~M.~Nomachi,~R.~Hazama\\
{\it Osaka University, Osaka, Japan}\\
\vspace{5mm}

F. Sergiampietri$^{1}$\\
{\it INFN,Pisa, Italy}\\
\vspace{5mm}

D. Naples \\
{\it University of Pittsburgh, Pittsburgh, PA}\\
\vspace{5mm} 

K. T. McDonald\\
{\it Princeton University, Princeton, NJ}\\
\vspace{5mm}

J.~Czakanski\\
{\it University of Silesia, Katowice, Poland}\\
\vspace{5mm}

A.~Godley,~S.R.~Mishra,~C.~Rosenfeld,~K.~Wu\\
{\it University of South Carolina, Columbia, SC}\\
\vspace{5mm}

J.T.~White\\
{\it Texas A\&M University, College Station, TX}\\
\vspace{5mm}

A. Mann, W.P.~Oliver, J. Schneps\\
{\it Tufts University, Boston, MA}\\
\vspace{5mm}

D. Cline, ~S.~Otwinowski,~Y.~Seo,~H.~Wang\\
{\it UCLA, Los Angeles, CA}\\ 
\vspace{5mm}

J.~Stepaniak\\
{\it Institute for Nuclear Studies, Warsaw, Poland}\\
\vspace{5mm}

W.~Dominik\\
{\it Warsaw University, Warsaw, Poland}\\
\vspace{5mm}

P.J.~Doe,~J.~Formaggio,~R.G.H.~Robertson,~J.F.~Wilkerson\\
{\it University of Washington, Seattle}\\
\vspace{5mm}

S. Menary\\
{\it York University, Toronto, Canada}\\
\end{center}

{\bf{$^{\dag}$ Contact person}}\\
{$^{1}$ Also UCLA, Los Angeles, CA} 
\newpage
\tableofcontents
\newpage
\newpage
\noindent
\vspace{1cm}

\begin{center}
{\LARGE \bf Executive summary}
\end{center}
\vspace{1cm}

\addcontentsline{toc}{chapter}{Executive summary}

Neutrinos have been a source of excitement lately.  While the neutrino mixing 
has been
expected, its pattern has taken us by surprise. The first evidence
of  physics beyond the standard model is here and more excitement
is likely on the way..

Fermilab has a unique opportunity to take a lead in the neutrino
physics in the next decade or so. Our recent major investments put us 
in an enviable position of having two intense neutrino beams with
energies spanning the range from $0.5~GeV$ to $20~GeV$. To exploit
fully our investment we need new experiment(s) matching the
physics potential of our neutrino beams.

Some of the most compelling questions can be addressed with a
Fermilab, long baseline, off-axis experiment.  A fortunate combination of
the NuMI off-axis beam spectrum and the emerging value of $\dmsq23$
makes the distances offered by the NuMI beam nearly optimal for the
$\nue$ appearance experiment, which is the key to the most interesting
questions:
\begin{itemize}
\item what is the pattern of neutrino masses? ``normal'' or ``inverted'' hierarchy?
\item what is the value of the mixing angle $\ssq2t13$ ?
\item is CP violated in the neutrino sector? 
\end{itemize}

We are very fortunate that a new powerful detector  technology
has been  developed recently. This technology, the Liquid Argon Time
Projection Chamber, is just waiting to be exploited. 
The Liquid Argon Time Projection Chamber is a relatively  inexpensive detector 
ideally matched to the need of neutrino experiments. It provides 
a bubble-chamber like view of interactions thus enabling  very efficient
detection and classification of neutrino interactions. The long R\&D effort
of the ICARUS collaboration culminated in the successful operation of a 300 ton 
detector. Using the technology demonstrated by the ICARUS group one can 
construct a wide variety of neutrino experiments.

The physics potential of  neutrino beams is limited by the statistics of the available
event sample which is proportional to a product of the beam intensity, the
detector mass and the detection efficiency. While it is important to maximize 
the beam intensity it is equally important to take full advantage of the 
available beam by maximizing the detector mass and the detection efficiency.
The high detection efficiency of a liquid argon detector makes its physics
potential equivalent to that of a conventional detector with the beam
power upgraded by the proton driver.

New, intense neutrino beams and advanced detection techniques have
rekindled interest in conventional neutrino scattering physics.
This physics is interesting in its own right, but improved  
knowledge in this area will also be necessary for precise measurements of the
neutrino oscillation parameters.

A very large, 50 kton class,  surface detector at the far location in the NuMI 
beam will enable sensitive measurements of the neutrino oscillation parameters.
Its superb spatial granularity may make it possible to 
detect supernovae and determine the resulting neutrino spectrum. It may even
detect protons decaying into kaons, should any of these events occur during the
lifetime of the experiment.

Small, 40 ton class, liquid argon detectors exposed to the MiniBOONE and NuMI 
beams at the close locations will provide a wealth of physics results,
thus filling a hole in our knowledge of neutrino physics at low energies.

Is the neutrino a Dirac or a Majorana particle? This fundamental question
awaits an experimental answer and  neutrino-less double beta decay
experiments are the way to provide it. These experiments require very
high background rejection and excellent energy resolution to fight
the background of a normal double beta decay. The Liquid Argon TPC provides
both, thus making it a very attractive candidate for the technology
of the next generation of experiments.

The Liquid Argon TPC appears to be a very versatile and very 
powerful technology applicable to a wide range of physics experiments. The
fundamentals of the experimental techniques have been established and, in 
principle, one can design small or large detectors as physics demands.
We feel, however, that more practical experience is necessary to make such
experimental proposals credible.

In the case of small detectors, the necessary R\&D is limited to 
technology transfer and gaining some practical experience. Credible
proposals for a very large detector require engineering studies of the
application of the fundamental technology on a very large scale. This
is especially important to confirm the credibility of  cost estimates
for such a detector. A double beta decay version of a liquid argon
detector also requires further studies of energy resolution at the
very low energy end.

Liquid Argon TPC detectors can greatly enhance the physics reach of the neutrino program
at Fermilab. They can also be of significant importance for neutrino programs
in future underground laboratories. Therefore we consider it necessary
to embark on a vigorous program of detector studies to establish credible
designs of  physics experiments. Fermilab, with its mix of technical resources
and a long tradition of liquid argon detectors, is a natural place for such 
studies.

The physics motivation and technical aspects of a Liquid Argon R\&D program are 
laid out in the following chapters:

\begin{itemize}

\item Chapter 1 describes the potential of a 50 kton off-axis detector for
studies of neutrino oscillations
\item Chapter 2 discusses the potential of a large surface detector to detect
supernova neutrinos and proton decay
\item Chapter 3 describes the physics potential of 40 ton detectors in the NuMI and 
MiniBOONE beams
\item Chapter 4 describes the fundamentals of the Liquid Argon TPC
 and the demonstration of its practical implementation by the ICARUS
collaboration
\item Chapter 5 describes a 50 kton off-axis detector and discusses technical
aspects and cost estimates

\item Chapter 6 describes a design of a 40 ton detector, suitable for the near
location at NuMI or MiniBOONE beams
\item Chapter 7 describes a possible application of liquid argon technology for
a neutrino-less double beta decay experiment
\item Chapter 8 outlines a possible program of technology transfer, engineering
 studies and possible improvements of the detector technology
\end{itemize}

\renewcommand{\thepage}{\arabic{chapter}-\arabic{page}}
\chapter{Neutrino Oscillations Studies with the Off-axis Liquid Argon Detector}
\section{Studies of Neutrino Oscillations: The Physics Case}

Neutrinos have mass and oscillate. There are two oscillation frequencies 
driven by the two mass differences of neutrino mass eigenstates: $\Delta m_{12}^2$  and
$\dmsq23$. Interference of these two processes is expected to produce measurable
CP violating effects.

The oscillation at the ``atmospheric'' mass difference $\dmsq23$ is a subject of 
studies with long baseline neutrino
experiments, the maximum of oscillations occurring at 
$E_{\nu}=\frac {2.54}{\pi } L \dmsq23$. To set the scale: taking the current estimate
of  $\dmsq23 = 2.5\times 10^{-3}~eV^2$ and a distance $L=850~km$ the oscillation
maximum occurs at $E_{\nu}=1.7~GeV$. 

Recent developments in neutrino physics have established that, contrary to
the expectations,  at least two neutrino mixing angles, $\ss2t12$ and 
$\sinsq2t23$ are very large. The latter is consistent with being the maximal,
hence raising a tantalizing possibility that some new, hereto unknown symmetry
of nature may be at work here. Precise measurement of the deviation of this angle
from its maximal value is, therefore, of great importance.

Neutrino oscillation experiments have established that neutrino masses exhibit
a pattern of a doublet of closely spaced masses, $m_1$ and $m_2$, whereas the third
mass is much further apart. A natural question arises what is the pattern of neutrino masses? 
Is it `normal', with $m_1 < m_2 \ll m_3$, or 'inverted', with $m_3 \ll m_1 < m_2$  hierarchy?
This question can be answered experimentally in the $\numutonue$ appearance experiment
with neutrino and antineutrino beams, as the the matter effects change the transition
rates differently. 

While 'solar' (including long baseline reactor experiment) and atmospheric neutrino experiments
have established that the electron neutrino is predominantly composed of $m_1$ and $m_2$
mass eigenstates there might be some admixture of the $\nue$ in the third state $m_3$.
The size of his admixture is known to be not very large,  $\ssq2t13 \le 0.12$, but its precise
value is of great interest. It has a significant power to discriminate between various
theoretical models of neutrino mixing but more importantly, its size determines the practical
observability of the CP violating effects in the long baseline experiments.

\begin{figure}[h!] 
\includegraphics{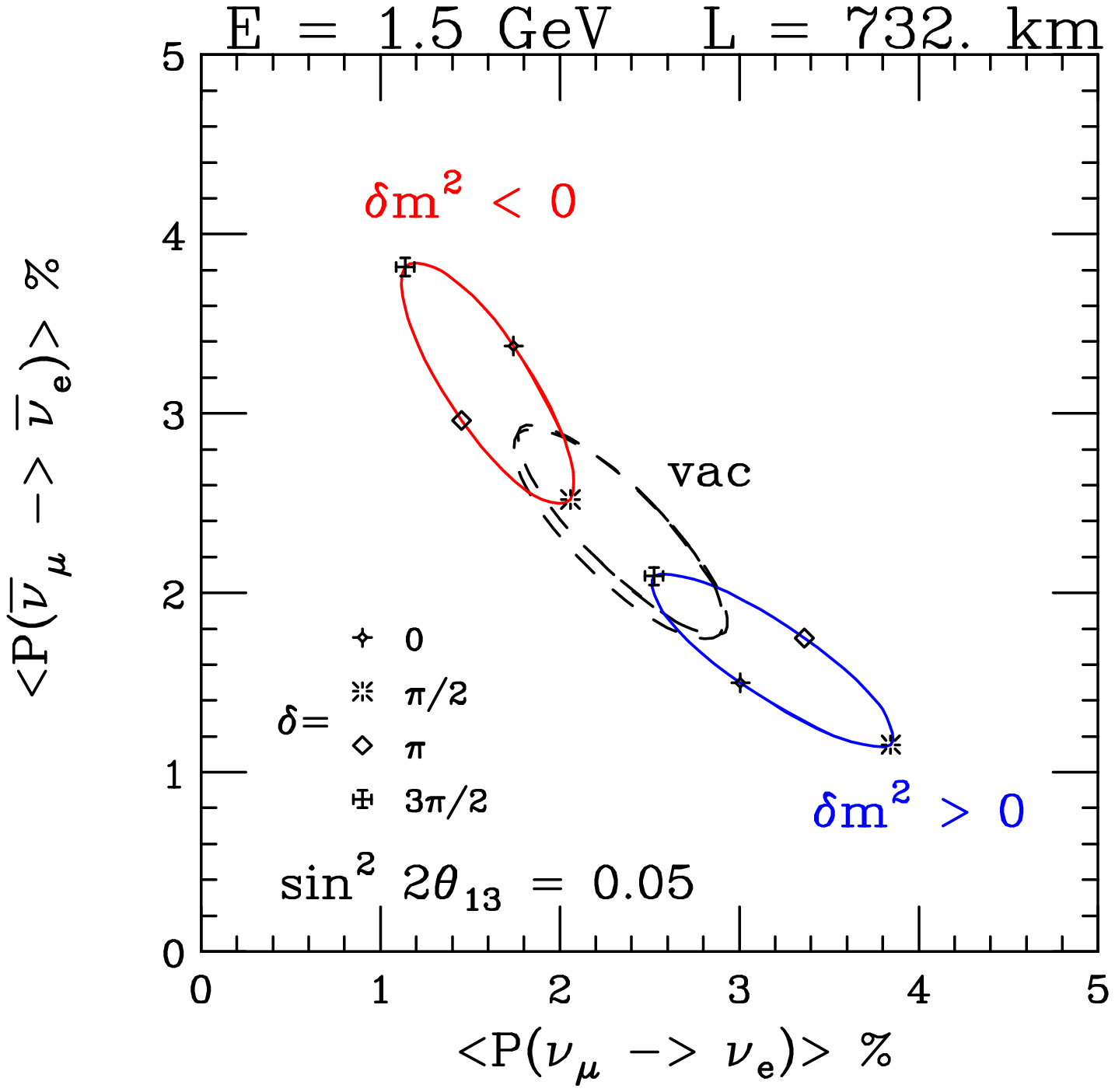}
\includegraphics{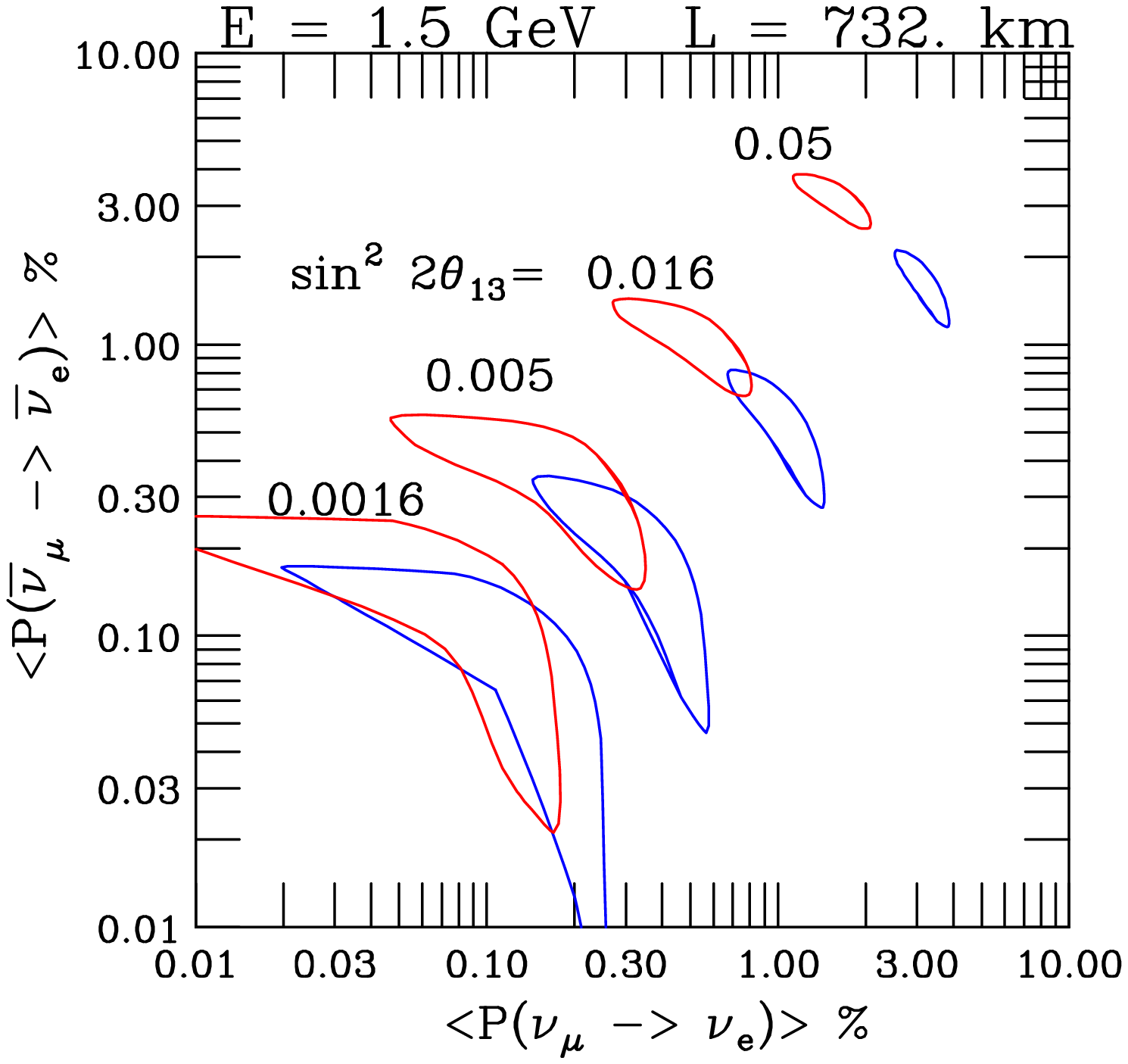}
\vspace{3.0in}
\caption{
The bi-probability plots $P(\nu_\mu \rightarrow \nu_e)$ versus 
$P(\bar{\nu}_\mu \rightarrow \bar{\nu}_e)$ assuming a constant matter
density of $\rho = 3.0 g cm^{-3}$ for an L/E of 500 km/GeV. 
The mixing parameters are fixed to be 
$|\Delta m^2_{31}| = 3 \times 10^{-3} eV^2$,
$\sin^2 2\theta_{23}=1.0$,
$\Delta m^2_{21} = +5 \times 10^{-5} eV^2$,
$\sin^2 2\theta_{12}=0.8$ with the labeled values of 
$\sin^2 2\theta_{13}$ and $\delta$.
}
\label{probsfig}
\end{figure}

It is quite fortunate that all of the emerging questions can be addressed by a single
experiment, that of $\numutonue$ oscillations. This is well illustrated on the 'bi-probability'
plots\cite{MinNun} shown in Fig.\ref{probsfig}. The experimental results can be expressed
as two numbers: $P_{\nu}$ and $P_{\nubar}$ being the probabilities of the $\numutonue$ transitions
in a given detector configuration for a neutrino and antineutrino beams respectively. 
A distance of the resulting data point is a measure of the mixing angle $\ssq2t13$. 
In a CP-conserving world and a vacuum experiment the result must lie on a diagonal. Matter
effects will displace the result to be below or above the diagonal for the ``normal'' and
``inverted'' hierarchy  respectively. The presence of th CP violating phase will move the 
result to in point on the ellipse,its exact location related to the value of the phase.

At the larger values of $\sin^2 2\theta_{13}$
the ellipses associated with the two possible mass hierarchies
separate in matter, whereas they are approximately degenerate in vacuum.
There is also a significant sensitivity to the CP violating phase, $\delta$.
It is the sensitivity to the sign of $\Delta m^2_{32}$  
and the CP violating phase in these
plots which allows for the determination of these parameters in 
a sufficiently accurate experiment.
For a single experiment there can be a degeneracy in the determined
parameters but this degeneracy can be broken by further experimentation.

\section{The NuMI Opportunity}

The NuMI neutrino beam was designed to address the ``atmospheric neutrino anomaly''.
By a fortunate coincidence its off-axis components\cite{BNL_off_axis,off_axis}
provide an unique opportunity to study the emerging questions in the neutrino
sector. The NuMI beam allows for experiments with a baseline between $700~km$
and $950~km$. The detector distance from the nominal beam axis determines 
the resulting beam energy and spectral shape. 

The current values of the oscillation parameters indicate that the oscillation
maximum, at the NuMI distances, occurs at the neutrino energies of the order of $2~GeV$.
This is very fortunate, as this energy corresponds to the decay angle of about $90^o$ 
in the center-of-mass of the decaying pion frame, hence the beam intensity is 
relatively high. The expected events rates, in the absence of oscillations, are
in the range of $100~events/kton/year$. Narrow energy spectrum helps to reduce the
backgrounds from possible neutral current interactions. 

The irreducible background for the $\nue$ oscillation experiment is due to the   $\nue$
component of the beam. The  $\nue$ flux is of the order of $0.5\%$ of the $\numu$ 
flux\cite{off_axis_loi} in the region of interest. In this region the  $\nue$ flux
is primarily produced by secondary muons decay, the same muons which accompany
the main $\numu$ component of the beam. It is expected hat the MINOS near detector,
as well as possible dedicated near detectors will determine this background
with an accuracy of few percent.

\section{Experimental Aspects of the $\nue$ Appearance}

The experimental challenge is to detect and identify charged current $\nue$ interactions
resulting from the $\numutonue$ oscillations. The signal events are characterized by
the presence of an electrons in the final state. The total energy of the final state
particles is a measure of the neutrino energy. A characteristic feature of the charged
current interactions, as opposed to the neutral current ones, is the balance of the 
transverse momentum, up to a small contribution from the Fermi motion of target nucleons.

The irreducible background of from the $\nue$ component of the beam has the same 
characteristics and it is indistinguishable from the signal in every detector.
 The only factor allowing to minimize this component of the background is the total
energy resolution, due to the fact the the signal events have their energy distribution
peaked sharply in the region of the maximum flux of the neutrino beam 
 whereas the background $\nue$ events have rather flat energy distribution.

The second contribution of the background is caused by events with no primary electron,
but the produced $\pizero$ or a photon conversion is mistakenly identified as an electron
in neutral or $\numu$ charged current interactions. The latter category is relatively 
suppressed as only the events with a muon not identified as such will be classified
as  potential signal candidates.

The ``fake'' electrons rejection is the primary experimental challenge of the experiment
and it determines the physics potential of the experiment. On one hand the presence of
the $\pizero$ background increases the background level, on the other hand a requirement to
keep it at tolerable levels leads to analysis cuts reducing the efficiency for the signal
detection.

The physics potential of an experiment (with a  given beam and at a given location) can be 
expressed in terms of
\begin{equation}
FOM = \frac{\varepsilon S}{\sqrt{\varepsilon (S+B_{beam})+B_{\pizero}}}
\end{equation} 
where S is the number of the events expected from the oscillations, $B_{beam}$ is a number of
CC $\nue$ interactions produced by the $\nue$ beam component, 
$\varepsilon$ is the
efficiency for the detection and identification  of $\nue$ CC interactions and $B_{\pizero}$
is the number of non-$\nue$ interactions classified as signal candidates under the cuts 
giving the $\varepsilon$ efficiency.

It is expected that the background levels will be determined with high accuracy and the physics
reach of the experiment will be limited by the statistical fluctuations of the otherwise well
known background. In such a situation $FOM \sim \sqrt{MtN_P}$ where $M$ is the detector mass, 
$t$ is the running time and $N_P$ is the number of protons delivered onto the neutrino target.

Several independent studies have shown that in the case of high granularity sampling detectors
the optimum of the physics potential is reached when the $\pizero$ induced background is of about
the same size as the intrinsic $\nue$ background and that such an analysis typically has
efficiency,$\varepsilon$, of about $30\%$.

A major improvement can be achieved with imaging detectors, like a liquid argon TPC,
which allow for nearly complete elimination of the $\pizero$ induced background and, at the
same time, high efficiency of the $\nue$ identification, of the order of $90\%$. 
The desired level of reduction of the $\pizero$ induced background is set by a requirement
 $\varepsilon B_{beam} \gg B_{\pizero}$. Further reduction contributes little to the $FOM$ as 
the statistical fluctuations of the irreducible background dominate.

The increase of the physics potential  of the imaging detector over the sampling calorimeter depends somewhat
on the physics scenario. If the osculation signal is large, hence 
$\varepsilon S \gg \varepsilon B_{beam}+B_{\pizero}$ than the liquid argon detector of a given mass
is equivalent to three times larger conventional detector, thanks to excellent electron 
identification efficiency. The ultimate sensitivity of the 
liquid argon detector, corresponding to the case $\varepsilon S \ll \varepsilon B_{beam}+B_{\pizero}$,
is equal to that of six times bigger conventional detector, or of the conventional detector
with a new proton driver-driven beam, thanks to an additional reduction of the background by a factor of two.

High detection and identification efficiency has an additional advantage of reducing
the systematic uncertainties. Corrections based on the event and detector simulations will
be very small and therefore not very much affected by a relatively poor knowledge of the
low energy neutrino physics.

\section{Particle Identification and Energy Resolution of The Liquid Argon Detector}

\begin{figure}[t!]
\begin{center}
\includegraphics[scale=0.3]{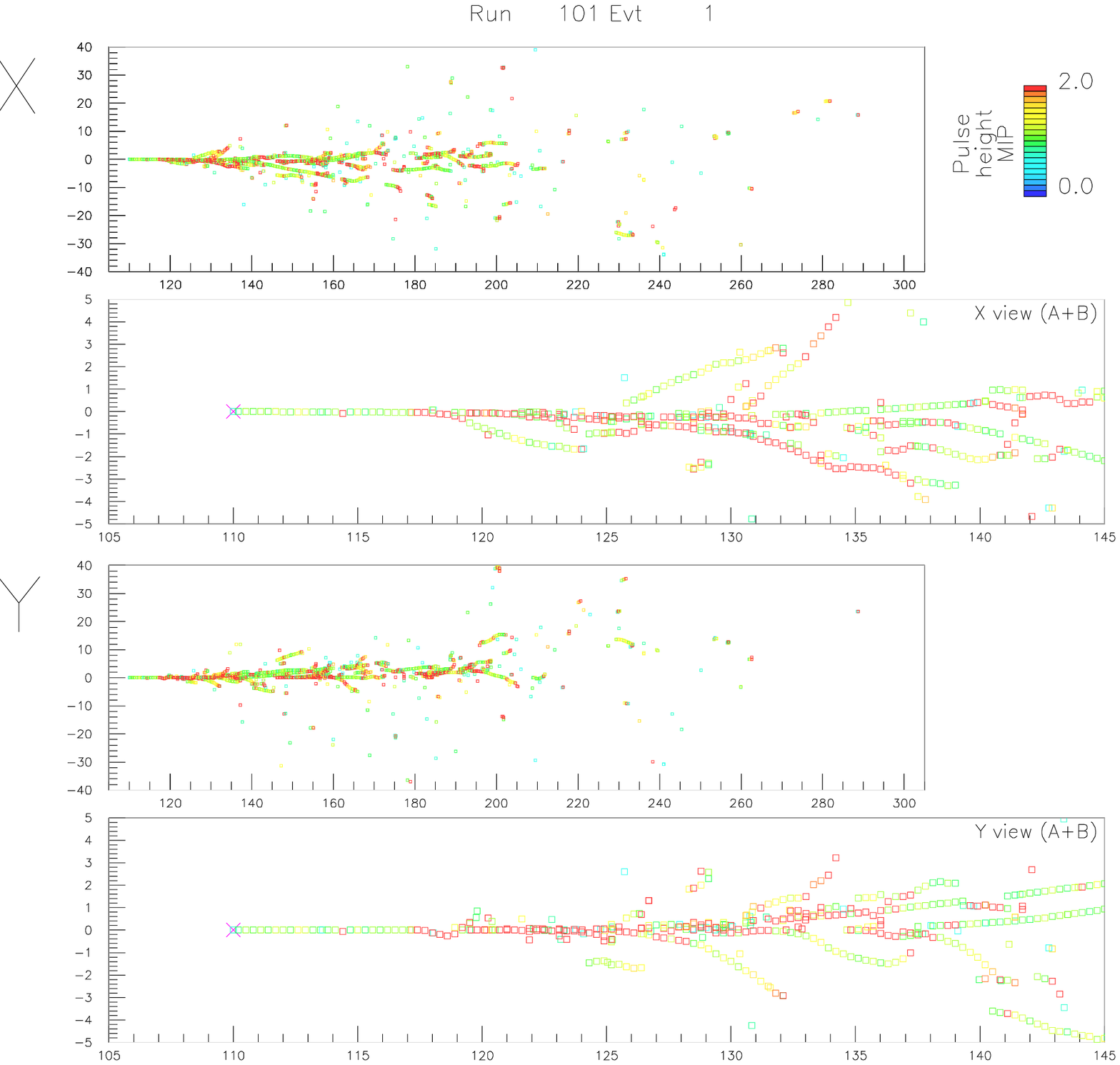}
\includegraphics[scale=0.3]{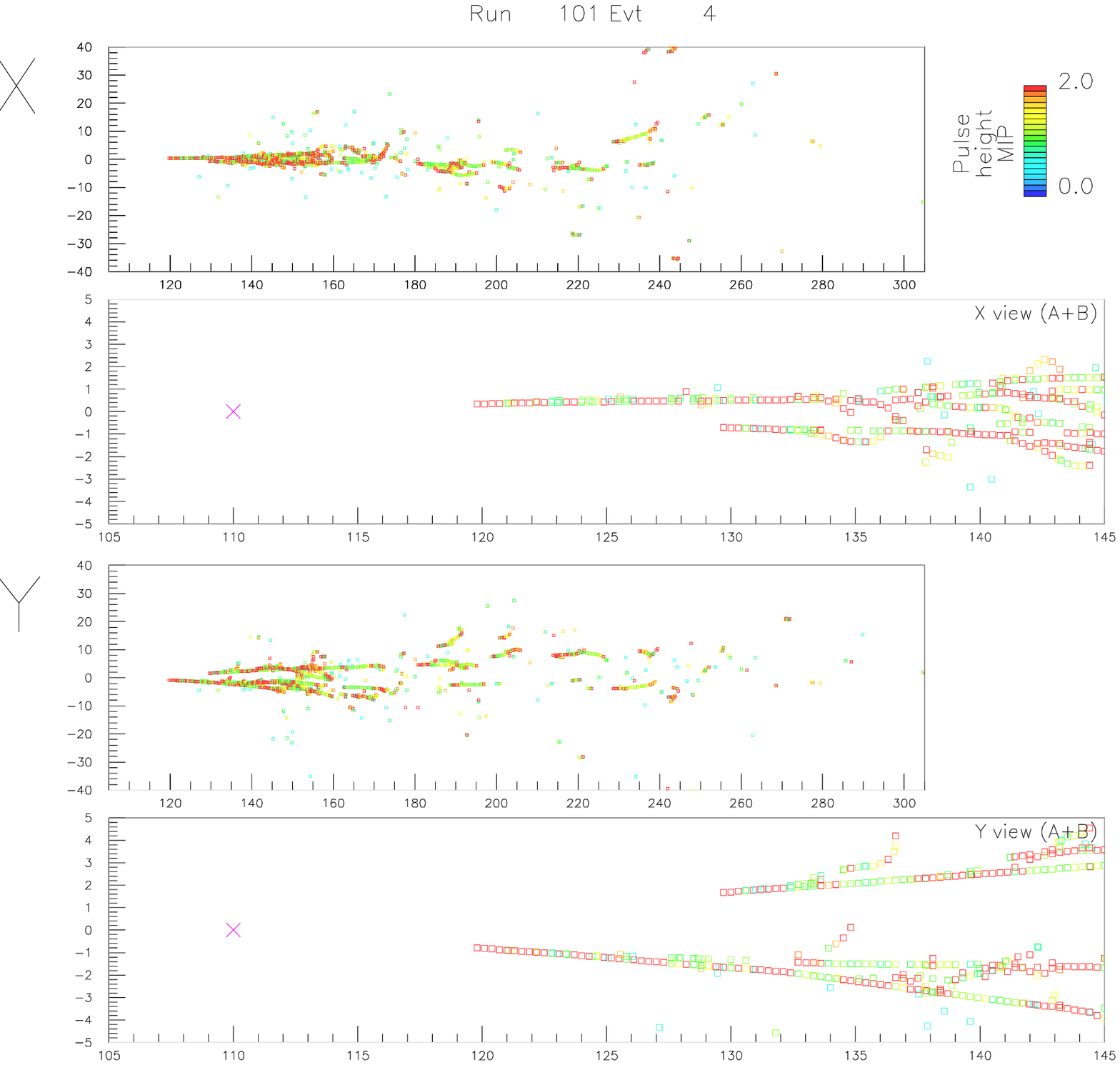}
\vspace{0.5cm}
\caption[]{Full GEANT3 simulation of a $1.5~GeV$ single electron (left) and $\pizero$ (right)
as detected in a liquid argon detector. Events are shown in windows $40~cm$ wide and $200~cm$ long,
followed by a blow-up of the vertex region $10~cm~\times~30~cm$. The particle origin is shown with
the pink cross. Color scheme  of the digitizations is adjusted such that a green ``hit'' indicates
a single MIP whereas a red ``hit'' indicates an energy deposit of two or more MIPS. It is important 
to notice that a 'tip' of a converting photos is red for a few $cm$ indicating a presence if
two overlapping particles.  }

\label{e_pi0}
\end{center}
\end{figure}

Imaging nature of the liquid argon detector combined with a very precise measurement of the ionization
density $dE/dx$ allows very efficient an unambiguous identification of stopping particles. 
Observation of the development of the electromagnetic cascade with very high spatial granularity
differentiates between 'electromagnetic' and 'non-electromagnetic' objects with high efficiency, 
which we assume to be $100\%$. The key to the $\nue$ appearance experiment is very efficient $e/\pizero$
separation. Fig.\ref{e_pi0} illustrates  some of the the basic features of the electron and
$\pizero$ events  recorded in the liquid agon detector.

The $\pizero$-induced background occurs in two distinct categories of interactions:
\begin{itemize}
\item (A) events with one or more charged particles produced at the interaction points. These
particles will define the interaction point with an accuracy of $few~mm$.
\item (B) events no no charged particles produced alongside $\pizero$. A coherent production
of $\pizero$'s off nuclei is a good example of such events
\end{itemize} 

Even ignoring the transverse shower information one can achieve a rejection factor
of $\pizero$'s vs single electrons of the order of 40 by demanding that the ionization
density over the first $2~cm$ of the track is consistent with a single track. Such a cut
will reject about $10\%$ of early showering electrons. For events in category (A) an additional
rejection factor of 15 is accomplished by a requirement that the electromagnetic shower
starts within $2~cm$ of the event vertex.

These two cuts are sufficient to reduce the number of $\pizero$-induced background well below that
of the intrinsic $\nue$ component of the beam. It is important to stress that there are many additional
rejection factors which more difficult to quantify without a complete reconstruction of the events
and which will provide redundancy and systematic checks of the background rejection procedure. They
include:
\begin{itemize}
\item neutral current events, with outgoing neutrino, will have a significant $p_t$ imbalance
with respect to the incoming beam direction.
\item distinct conversion points of two separate photons will be often visible. Owing to an excellent
energy resolution an invariant mass of the $\pizero$ can be reconstructed.
\end {itemize}

In th following we will, therefore, assume that the electron identification a the level of $90\%$
with a $\pizero$-induced background negligible with respect to the intrinsic $\nue$ component
of the beam can be attained.

The liquid argon detector is, at the same time, a total absorption calorimeter and has a very good energy 
resolution. Electromagnetic shower energy can be measured with a resolution better that $1\%$. Stopping
hadrons will have their energy determined to a few \%, whereas an energy of an hadron-initiated
cascade will be measured with an typical accuracy $\Delta E/E = 0.35/\sqrt{E}$.
Implications for the accuracy of determination of the initial neutrino energy are less obvious:
the energy resolution of the total observed charge as a measure of the incoming energy is limited
by nuclear absorption effects and by kinematical factors like rest masses of the produced particles.
It is of great help, in the energy regime of interest, that owing to a relative simplicity of the neutrino 
interactions  all of the final state particles can be detected and their rest mass included in the neutrino
energy determination. With this correction the incoming neutrino energy can be reconstructed with
an accuracy $\Delta E/E \sim 10\%$. For some selected categories of events (like quasi-elastic scattering
with lepton + proton in a final state) the neutrino energy resolution of the order of $1-2\%$ can be achieved.

\section{Physics Potential of The Liquid Argon Detector}

Physics potential of a specific off-axis experiment depends somewhat on the detector location.
We have not performed the optimization process, yet but we use as an example a 50 kton detector
located at the distance of $820~km$ from Fermilab and $11~km$ off-axis. We assume that $85\%$ of 
the detector volume will be used in the experiment. We will assume that the efficiency to identify
the $\nue$ CC interaction is $90\%$ and that the background is dominated by the intrinsic $\nue$
component of the beam, which is well constrained by measurements at the near detector.

\begin{table}
\begin{center}
\begin{tabular}{|l|r|r|}
\hline
    & Neutrino & Antineutrino  \\ \hline
Number of $\numu$ CC events, no osc.  &  18,906 &  5,714  \\ \hline
Background                             &      56 &     23 \\ \hline
Signal ($P_{osc}=0.05$ )                 &     527 &    158 \\ \hline
$P_{limit}$ (3$\sigma$)                   &     0.002&   0.004  \\ \hline
$P_{limit}$ ($90\%~CL$)                   &     0.0012 &  0.0025  \\ \hline

\end{tabular}
\end{center}
\caption{Event rates and attainable limits on the oscillation probability 
for 5 years run of 50 kton detector with medium energy
NuMI beam and $4\times 10^{20} POT/year$.}
\label{rates}
\end{table}

The experiment will measure the probability of $\numutonue$ transition.
The expected signal and background  rates as well as the sensitivity of the experiment
is summarized in Table.\ref{rates}.
 
Comparison of the attainable levels of probability with the Fig.\ref{probsfig} gives
an indication of the sensitivity of the experiment in terms of the physically interesting
parameters: mixing angle, mass hierarchy and and CP phase $\delta$. It is quite obvious
that there is no unambiguous answer to a question: what is the sensitivity of the
experiment to the $\ssq2t13$?
 
\begin{figure}[t!]
\begin{center}
\includegraphics[scale=0.3]{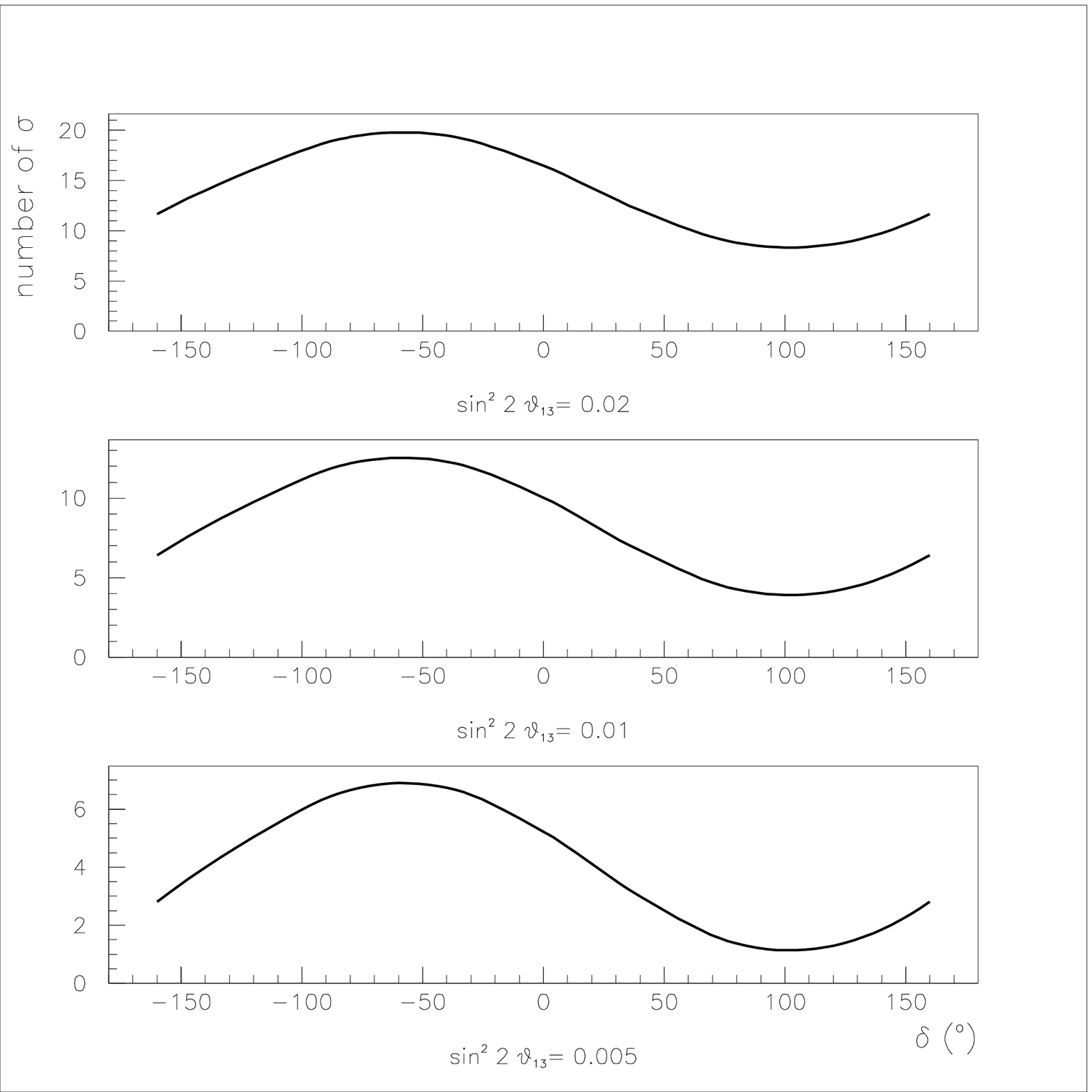}
\includegraphics[scale=0.3]{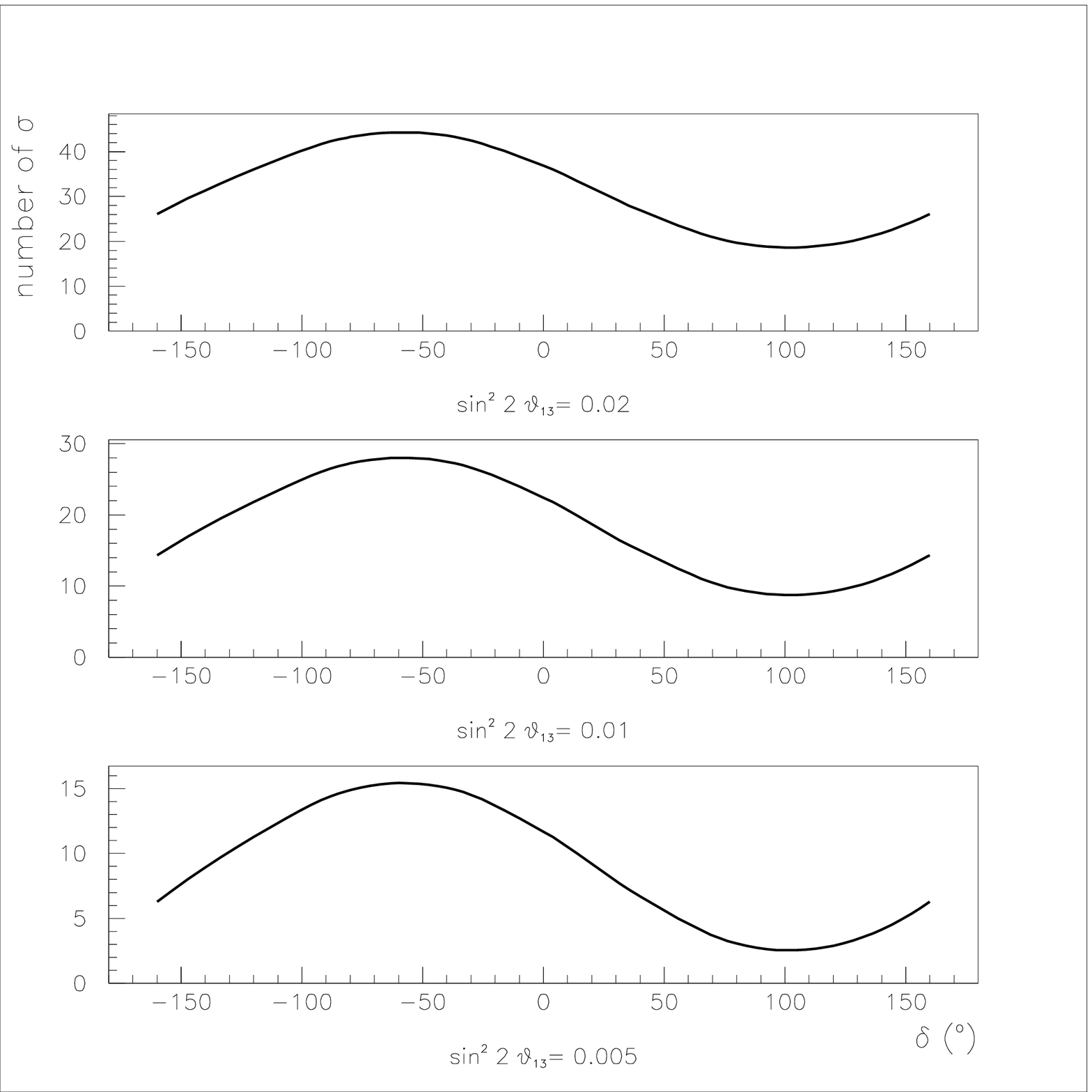}
\vspace{0.5cm}
\caption[]{Significance of the observed excess of $\nue$ CC events in 6 years run of a
50 kton liquid argon detector in the NuMI beam enhanced with the Proton Driver as a 
function of the CP phase $\delta$
for different mixing angles $\th13$. Left graph is for the inverted hierarchy, right one for the normal hierarchy\cite{Olga_Mena_1}.  }

\label{discovery}
\end{center}
\end{figure}

The ultimate sensitivity can be achieved with the increased NuMI beam intensity
with the proposed Proton Driver. Fig.\ref{discovery} shows the expected significance of the observed excess of 
the $\nue$ CC events for different mixing angles and for different mass hierarchies,
for different mixing angles and for different
values of the CP phase. It shows that the ultimate off-axis experiment can reach several standard 
deviations discovery even in the physics scenarios  beyond the reach of the proposed reactor 
experiments.

\section{Program of the Neutrino Oscillation Studies with FLARE at NuMI}

A liquid argon detector in a nominal NuMI beam will enable a sensitive measurement
of the $\numutonue$ transition probability for neutrino and antineutrino beam.

Can such a single experiment provide an unique answer to all the questions mentioned earlier?
What will be required to sort out possible ambiguities in the interpretation of the
experimental results in terms of physically interesting parameters? We cannot answer these
questions now, but we argue that the proposed detector offers a long range scientific program 
aiming at the ultimate determination of the parameters of the neutrino oscillations. 

Fig.\ref{probsfig} shows that possible results consistent with the oscillation interpretation
lie in two bands corresponding to two possible neutrino mass hierarchies: normal (blue ellipses) 
and inverted (red ellipses). These bands are well separated, in  bi-probability space, when
the mixing angles is relatively large and they merge when the mixing angle is small.

There four possible classes out outcomes of the off-axis experiments and we will examine
them in turns and we will argue that a liquid argon detector offers the best experimental
possibility in each case:

\begin{enumerate}

\item the resulting $P_{\nu}$ and $P_{\nubar}$ lie outside the region expected in the 
oscillation scenario. 

Such an unexpected result would not be the first failure of the 
theoretical expectations. It would indicate that 'something else' is happening. The purity
of the sample and the completeness of the event information offered by the liquid argon
detector might be of crucial importance in establishing what is going on.

\item the result is at one of the outer edges of the physics region. 

This would be  very kind of the Nature. In such a case the mass hierarchy is clearly indicated,
maximal CP violation is established, the CP phase is quite well known, mixing angle is
well determined. The physics interpretation is unambiguous and the further increase of  
statistics will reduce the errors on the physics parameters. The advantage of the 
liquid argon detector here lies in its high detection efficiency, making it equivalent to
much larger conventional detector or much brighter neutrino beam.
  
\item the result is somewhere in the middle of the allowed region, where two bands overlap.

This may well be the most likely outcome. Depending on the exact location of the result there 
will be some level of ambiguities in the physics interpretation of the experimental results.
We illustrate this case by taking an example of the possible result: let $P_\nu=0.0167$ and
$P_{\nubar}=0.0173$. From Table.\ref{rates} we find that they will be measured with the relative
accuracy of $8\%$ and $14\%$ respectively after 10 year run in the nominal NuMI beam, or in one 
year exposure in the Proton Driver era. Fig.\ref{amb}(left) shows that there are four different 
combinations of physics parameters which are consistent with such a measurement. Further reduction
of the statistical sharpness the allowed regions of phase space but it will not resolve
the ambiguities.

\begin{figure}[t!]
\begin{center}
\includegraphics[scale=0.3]{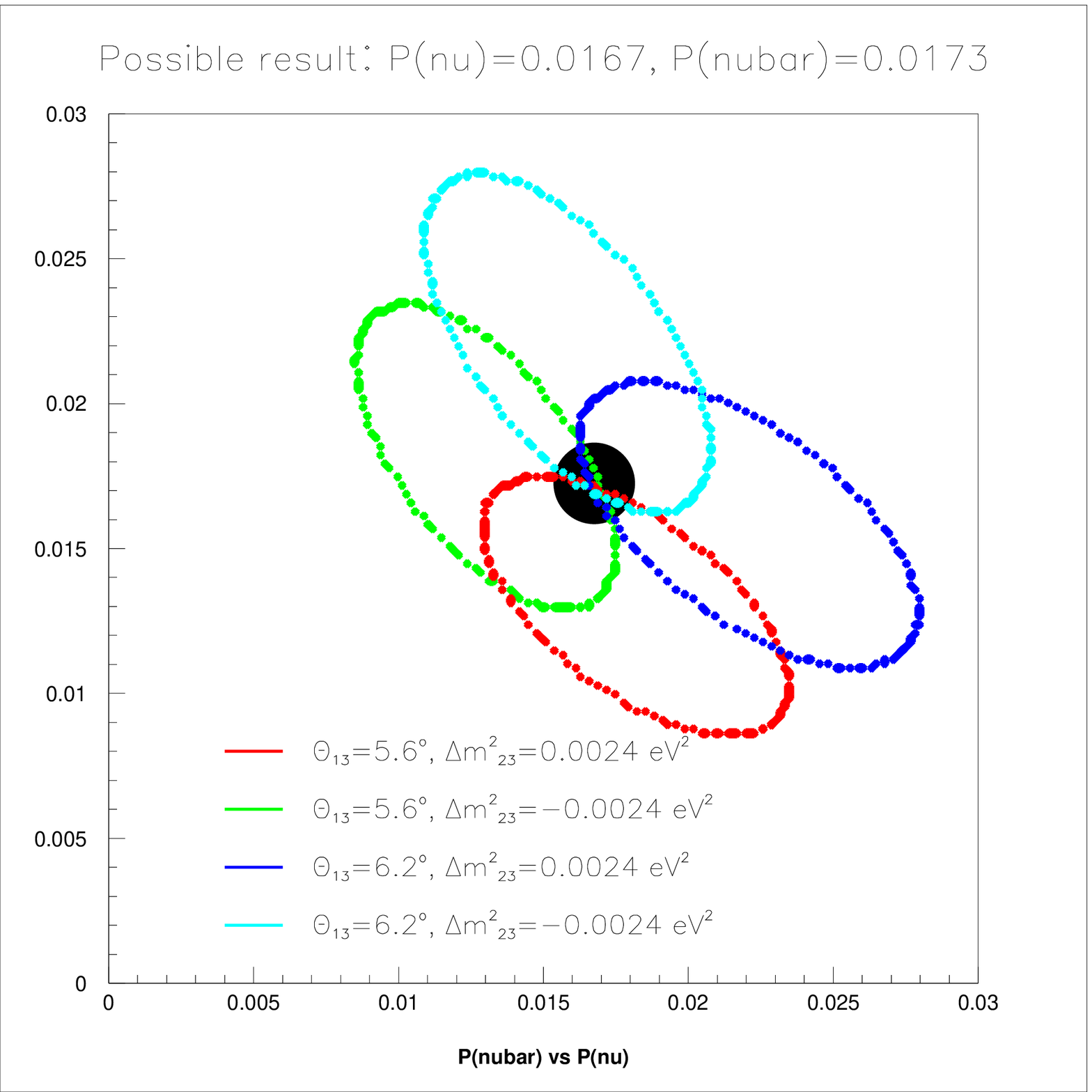}
\includegraphics[scale=0.3]{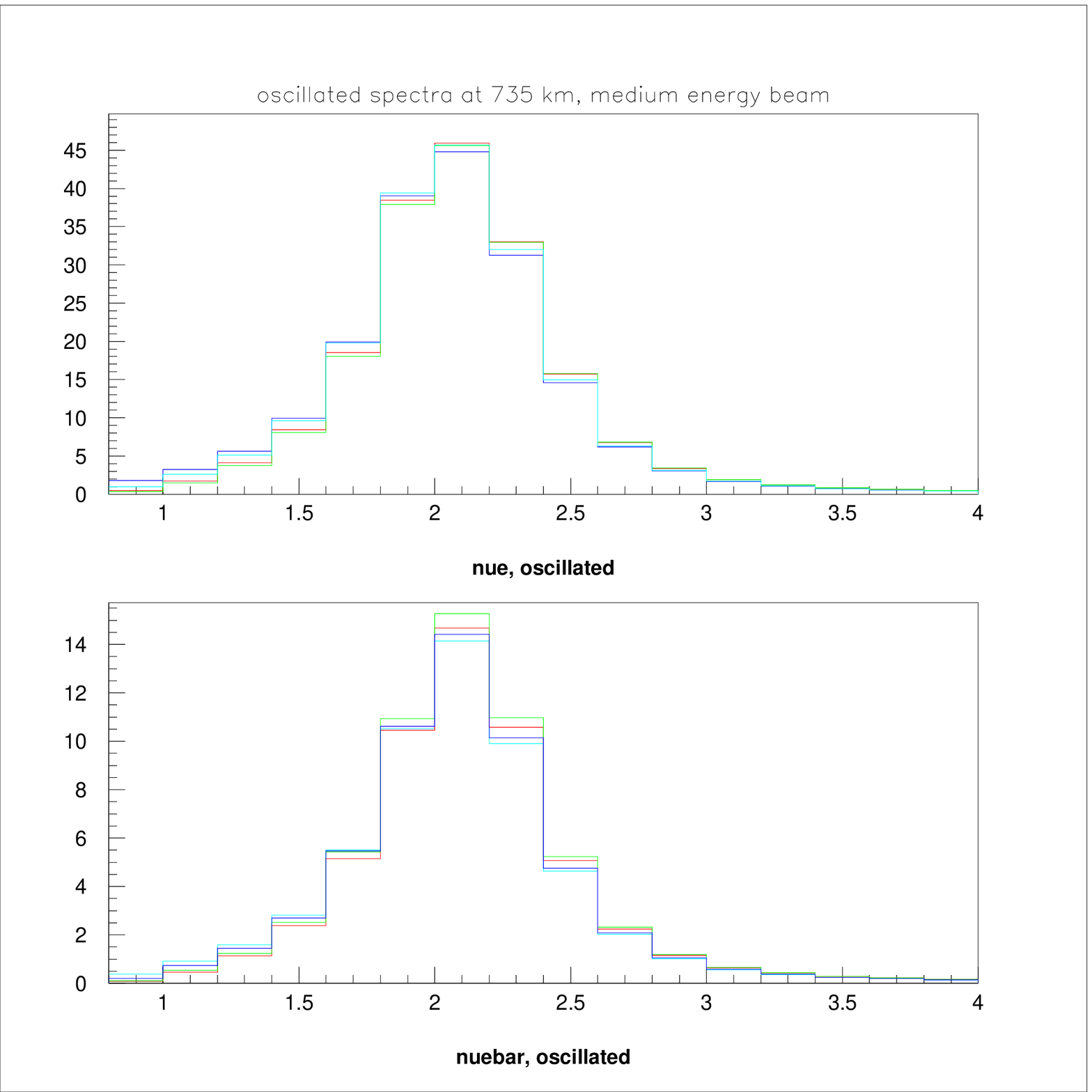}
\vspace{0.5cm}
\caption[]{Left: Values of $P_{\nu}$ and $P_{\nubar}$ expected for different values 
of the CP phase $\delta$ and two mixing angles for both mass hierarchies. Right:
Spectra of $\nue$ CC events produced in four different physics scenarios
consistent with the selected values of  $P_{\nu}$ and $P_{\nubar}$.
Colors of histograms correspond to the sets of parameters indicated on the
left graph. Top graph is the
neutrino beam case, bottom is the antineutrino beam case.  }
\label{amb}
\end{center}
\end{figure}

As the Fig.\ref{amb} (right) shows even a precise measurement of the energy spectra
of oscillated $\nue$ CC events does not help very much: they are nearly identical in 
all four cases for the NuMI beam conditions. The degeneracy can be broken and an unique
interpretation can be achieved by combination of these results with the results of an
experiment carried under different conditions. A longer baseline and accordingly bigger
mass effects are of particular help.  Fig.\ref{BNL}(left) shows an example of the  $\nue$ CC events
detected in a NuMI-like low energy beam aimed at the NuMI off-axis detector from Brookhaven
for the four, previously ambiguous physics scenarios. Ambiguities will be resolved and the 
parameters neutrino oscillation parameters determined with the accuracy given by he available
statistics. Fig.\ref{BNL} (right) shows that the NuMI/BNL beam combination is a very powerful
tool to determine the mass hierarchy. The observed rate of $\nue$ appearance are different
because of difference in the matter effects which, in turn, depend on the mass hierarchy.    

\begin{figure}[t!]
\begin{center}
\includegraphics[scale=0.25]{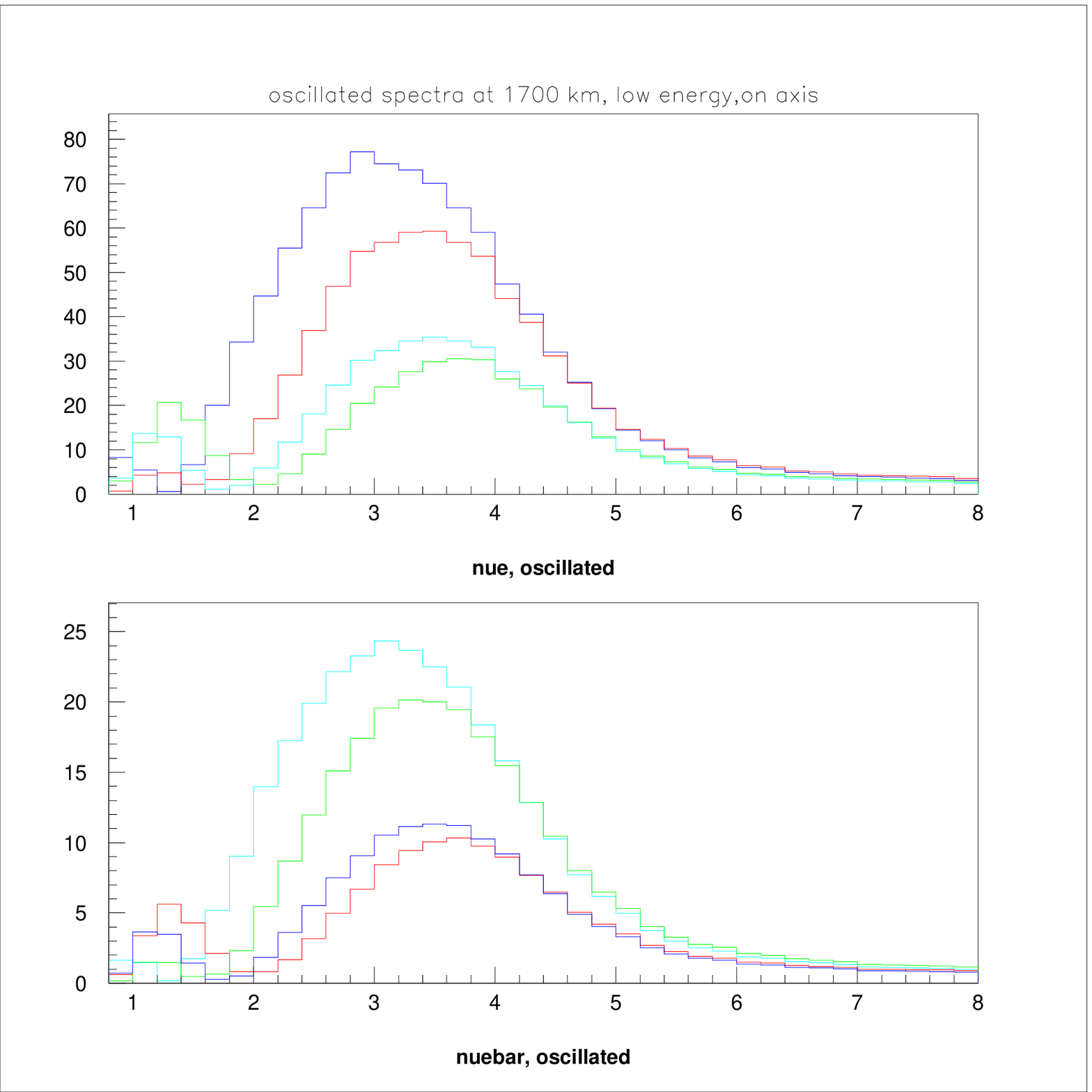}
\includegraphics[scale=0.5]{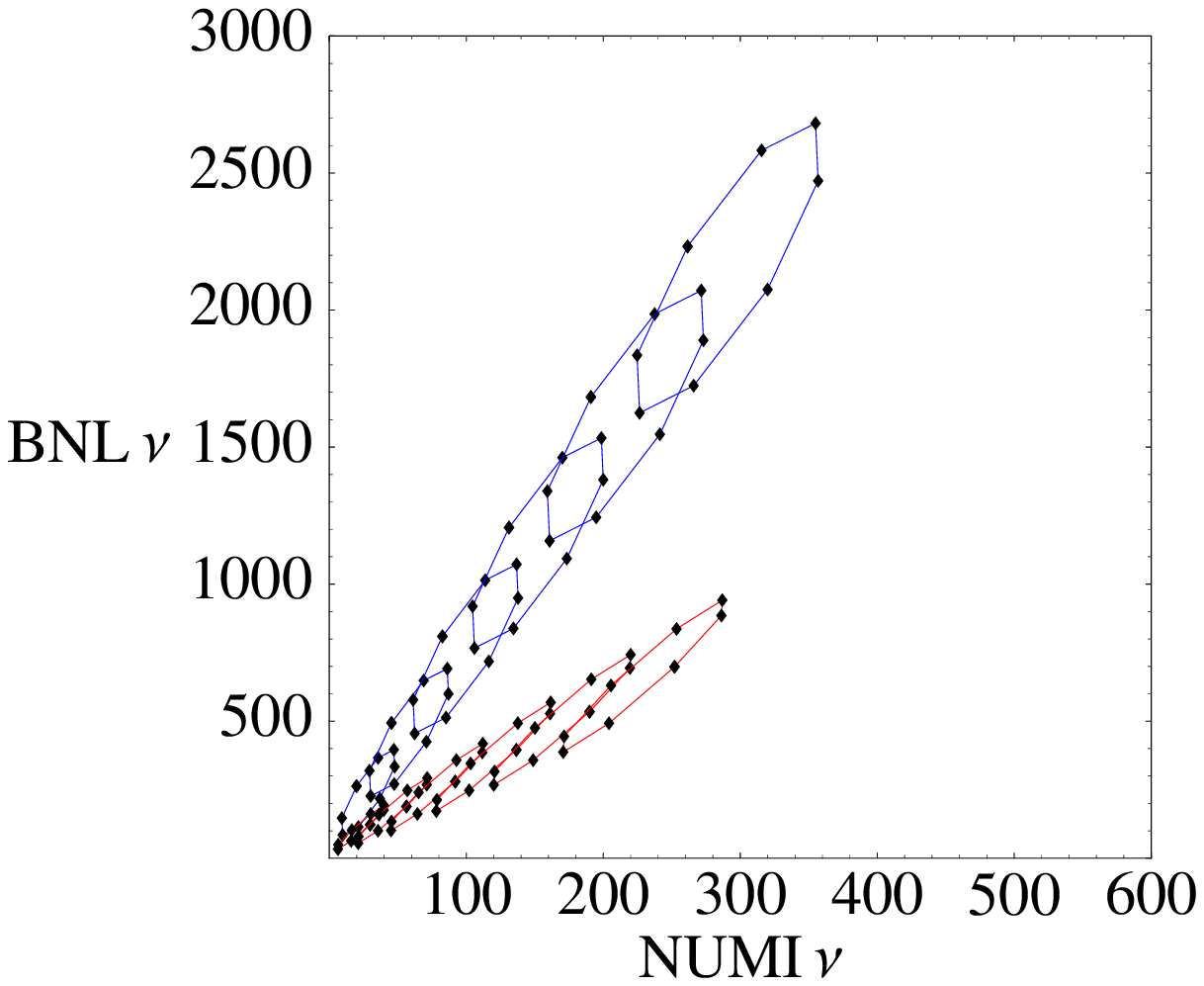}
\vspace{0.5cm}
\caption[]{Left: Spectra of the $\nue$ CC events detected in the low energy beam produced
at Brookhaven and directed towards an Off-axis liquid argon detector. Four different colors 
correspond to the four ambiguous (in the NuMI beam case) physics scenarios. Colors of
histograms correspond to the physics parameters as in Fig.\ref{amb}. Right: comparison of the
observed   $\nue$ appearance rates in the NUMI and BNL beams. Different ellipses correspond to different values of the mixing angle\cite{Mena_2}.  }
\label{BNL}
\end{center}
\end{figure} 

In this particular case the advantage of the liquid argon detector, beyond the usual efficiency
argument, is its isotropy of the good detection efficiency. This is important as the BNL beam
comes from considerably different direction and a conventional planar detector would not
be quite optimal. A good energy resolution with no long tails of the resolution function
and very good rejection of $\pizero$-induced background is very important for the wide-band BNL 
beam.
 	
\item the experimental result is in vicinity of the origin of the coordinate system

A very small or none signal of $\nue$ appearance signal is detected. The mixing
angle  $\ssq2t13$ is shown to be very small or a very tight limit is set. There is no information
on the mass hierarchy, nor on the CP violation. In this case the limit obtained by a liquid argon 
detector is equivalent to that of six time larger conventional detector (or six times more
intense neutrino beam), the reduction of background (relative to the conventional detector) 
contribution an additional factor of two.

\end{enumerate}

A null result of the several-year long appearance experiment would produce a very tight 
limit on the mixing angle, but it would be somewhat disappointing nevertheless given
the scale of invested effort. It is therefore of great importance to examine other possible 
uses of such a detector. 
\chapter{Physics Potential of a Surface Liquid Argon Detector}
 
The primary motivation for an of-axis detector is the study of neutrino oscillations
with the NuMI beam. Given the magnitude of the investment in the detector on one hand
and a possibility or relatively un-eventful run if the mixing angle is very small it
is very important to examine other possible physics topics which can be addressed with
such a detector.

\section{Free running surface detector operation}

To study neutrino beam related interactions it is sufficient to record the 
waveforms of the wires signals in the time frames corresponding to the beam spill.
The beam arrival time establishes the reference time for the measurement of the
drift time. The duration of the neutrino extraction, $10~{\mu}sec$ introduces 
an error on the absolute position of the event of the order of $1.5~cm$ but this is
inconsequential for the reconstruction of the interaction. 

In order to be able to address other physics topics, like a search for rare phenomena,
it is necessary to read out the detector continuously. This leads to several potential 
problems which are addressed in this chapter:
\begin{itemize}
\item large volume of data
\item lack of the time reference, $T_{0}$
\item cosmic rays induced backgrounds
\end{itemize} 

It is important to point out that the
data recorded in T300 module in Pavia were taken at the surface, thus they provide a very 
important check on the feasibility of the surface detectors.

\subsection{Data rates and data acquisition}

It is expected that the front-end electronics will provide a significant data reduction by
performing cluster finding and baseline subtraction. The amount of data passed to the data 
acquisition system  may contain some raw digitization in the region-of-interest (ROI)
in the early stages of the experiment, but it is expected that after the initial 
debugging phase of the experiment only the 'hits' corresponding to the physical signals will be
recorded. The data rates will be dominated by cosmic muons. Large electromagnetic showers or
hadron-initiated cascades occur as well, but their rate makes them to  contribute little to the
average data rate.  The effective area of the detector is of the order of $2,000~m^{2}$ hence
the average rate of cosmic muons is about $200~kHz$. Majority of muons will range out in 
the upper part of the detector. Ranging out and angular distribution of the cosmic rays in 
conjunction with almost vertical wires geometry  will lead to relatively small number
of wires recording signals. To set a safe upper limit we assume that the average number of
hit wires is less than 2000. Assuming 4 bytes per digitization it leads to the expected 
data rate of $1.6~Gbytes/sec$, well within capabilities of a relatively modest data acquisition
system comprised of few hundred of PC's. 

While the instantenous data rate does not present a major problem, the overall data size
is much more challenging: $100~ Tbytes/day$ and $30~pbytes/year$. Data storage and analysis 
systems with such capabilities are likely to be relatively standard in the near future, but
they will nevertheless require substantial investment.

It is very likely, however, that the data volume problem will be greatly reduced, by
a factor of thousand or more. Vast majority of the data will be associated with cosmic
muon tracks. They can be easily recognized by the data acquisition system and subsequently
removed from the data stream. Even in the case where such data will be of physics interest
it is very likely that a small fraction like 0.001 of them will be sufficient.
 
\subsection{What is a $T_0$??}

The distance of particle trajectory from the wire plane is determined by the drift distance,
which is the difference of the arrival time and the 'production' time. The former
is recorded by the electronics. The latter one is provided by th beam gate for the 
accelerator-oriented experiment. What is the $T_{0}$ for an object recorded asynchronously,
like supernova neutrino. 

In the large detector there will be several 'tracks' drifting the sensitive volumes at
the same time, hence there is no global $T_{0}$ for an 'event' or a time frame. Instead,
the reference time is a property of the 'object' in question. How can one determine what is 
the appropriate $T_{0}$ for a given set of digitizations forming a trajectory?

Before answering this question let's examine a consequence of applying an incorrect value
of $T_{0}$ in the analysis of a given trajectory. Too early $T_{0}$ used  will lead to an
assignment of a too large of a distance from the wire plane to the given track. If the track
in question crosses the cathode plane then the points in vicinity of the cathode will be 
assigned unphysically large distance, thus indicating that the  $T_{0}$ value used was 
incorrect. In a similar manner, if too late  $T_{0}$ is used and the track crossed the wire
plane it will b assigned negative distance. Tracks crossing the cathode plane or the wire 
plane do determine their own $T_{0}$.

 Tracks segments fully contained within one drift volume  will have a range of possible 
$T_{0}$ values. The correct one can be determined by using the observed ionization of delta 
electrons along the trajectory. The energy of the electrons can be determined from their 
range, which is translationally invariant, hence independent of the assumed $T_{0}$, hence
the produced signal is known for them. The drift distance, and hence the $T_{0}$ for the
object in question,  can be determined from the measured 
signal corresponding to the delta ray as the attenuation is known from the continuously
monitored electron lifetime.

\section{Cosmic rays: background or signal?}

High rate of cosmic muons leading to a huge data volume is likely to be a nuisance.
Almost all of the can be easily recognized and discarded by the data acquisition system.
This is a very likely scenario. 

We should note, however that this detector will produce unprecedented volume and quality
of data on the cosmic ray fluxes. Large area covered, high spatial resolution, good
particle identification will provide a new window of the cosmic ray fluxes. Composition 
and spectra of cascades, correlations between electromagnetic and hadronic components, 
other correlations can be studied. It is not clear  if such a dedicated experiment is
worth the effort, but why not look at this data before throwing them away?.

\section{Supernova Neutrinos and Physics Potential}

The core collapse of a massive star is an abundant source of
neutrinos and antineutrinos of all flavors, with energies in
the tens of MeV range, and which are emitted over a timescale of
tens of seconds.  A core collapse neutrino signal
was detected for SN1987A in the Large Magellanic Cloud, 55 kpc away, by
the Kamiokande-II and IMB water Cherenkov detectors.  Several
neutrino detectors online now are sensitive~\cite{Scholberg:2000ps}.

The neutrinos from a core collapse in our Galaxy will bring a wealth
of information.  Note only will they illuminate the details of the
core collapse mechanism via their time, flavor and energy structure,
they will also provide a tremendously bright source for fundamental
physics.  Although the potential for kinematic mass limits (which for
SN1987A were among the best at the time) is now exceeded by laboratory
experiments, again the energy, flavor and time structure of the
neutrino burst may provide information on fundamental neutrino
parameters.  For instance, flavor transitions in the stellar core due
to matter effects may occur, depending on the value of $\theta_{13}$,
and whether the neutrino mass hierarchy is normal or inverted can
dramatically change the nature of the supernova signal. These
signatures have been explored in several
references~\cite{Bueno:2003ei,Gil-Botella:2003sz, Gil-Botella:2004bv}.
The signatures are somewhat core collapse model-dependent, but there
are some relatively robust features of the neutrino signal, on which
one can base oscillation studies e.g. the hierarchy of energies for
different flavors
($\bar{E}_{\nu_{\mu,\tau}}>\bar{E}_{\bar{\nu}_{e}}>\bar{E}_{\nu_{e}}$).
Better understanding of the collapse mechanism via the
neutrino signal feeds back to neutrino physics (and vice versa).

Supernova neutrino detectors require energy and direction sensitivity,
and \textit{good flavor sensitivity and tagging ability} is essential
for extracting physics.  Because the largest detectors currently
online have a primary sensitivity to $\bar{\nu}_e$ via inverse beta
decay, sensitivity to flavors other than $\bar{\nu}_e$ is especially
valuable.  In particular, because an interesting feature-- a sharp
burst of ${\nu}_e$'s, due to ``shock breakout'' -- is expected near
the onset of the signal, ${\nu}_e$ sensitivity is especially
desirable.

Finally, there is the potential for an early alert to astronomers,
since the photons associated with the disruption of the star (the
supernova itself) will arrive later than the neutrino signal by hours
or perhaps longer, dependent on the nature of the stellar envelope.
For an early alert, pointing capability, i.e.  sensitivity to neutrino
interactions which preserve directional information, is highly
desirable.

The expected frequency of stellar collapses in our galaxy is about one
per thirty years, which is often enough to give one a reasonable hope
of seeing one on the timescale of a large neutrino experiment, yet
rare enough to motivate special care to extract maximum information
from an event.

\subsection{Supernova Neutrino Signal in Liquid Argon}

The supernova neutrino signal in liquid argon has been studied in some
detail in
references~\cite{Bueno:2003ei,Gil-Botella:2003sz,Gil-Botella:2004bv},
in the context of an underground detector.  There are three main
classes of relevant neutrino interactions: elastic scattering, charged
current, and neutral current.

\begin{itemize}

\item Elastic Scattering

Elastic scattering of neutrinos off atomic electrons,

\begin{equation}
\nu_x~+~e^-~\rightarrow~\nu_x~+~e^-,
\end{equation}
has both CC and NC components, and can provide pointing information.
The signature in liquid argon is a single electron track.

\item Charged Current Interactions

Neutrinos and antineutrinos can interact
via quasi-elastic charged current reactions with neutrons and protons
in nuclei:
\begin{equation}
\nu_e~+^{40}Ar~\rightarrow~e^-~+~^{40}K^{*},
\end{equation}
with a threshold of 1.5~MeV
and 
\begin{equation}
\bar{\nu}_e~+^{40}Ar~\rightarrow~e^+~+~^{40}Cl^{*},
\end{equation}

with a threshold of 7.48~MeV.  At high energy, inelastic
charged-current processes may yield neutrons or other particles as
well; however, in the tens of MeV regime relevant for most of the
supernova neutrino signal, the contribution is relatively small.
CC interactions may be identified via the $e^{\pm}$ track, in
coincidence with de-excitation gamma rays from the resulting nucleus.

\item Neutral Current Interactions

The neutral current excitation of argon,

\begin{equation}
\nu_x~+^{40}Ar~\rightarrow~\nu_x~+^{40}Ar^{*},
\end{equation}

for both neutrinos and antineutrinos, can be detected via its
de-excitation gammas.  Neutrons or other particles may also be
produced at higher energies.

\end{itemize}

The number of interactions scales simply with mass, and as
1/distance$^2$ Table~\ref{tab:snevents}, based on
ref~\cite{Gil-Botella:2004bv}, shows expected numbers of events in a
50 kt detector for these channels, for a core collapse at 10~kpc,
assuming no oscillation effects.

\begin{table}[t]
\begin{centering}
\begin{tabular}{||c|c||}\hline\hline
Channel & Number of events at 10 kpc \\ \hline\hline
Elastic scattering & 665\\ \hline \hline
Charged current $\nu_e$ & 3120\\ \hline \hline
Charged current $\bar{\nu}_e$ & 270  \\ \hline \hline
Neutral current $\nu_x$ & 15220  \\ \hline \hline
\end{tabular}
\caption{Numbers of supernova neutrino events expected in a 50~kt liquid
argon detector.}\label{tab:snevents}
\end{centering}
\end{table}

The relative numbers of events in each channel may change dramatically
due to flavor transitions in the stellar core, depending on the values
of neutrino mixing parameters.  If one assumes that these channels can
be tagged effectively, and the parameters are favorable, one may have
sensitivity to both $\theta_{13}$ and the mass hierarchy.
References~\cite{Bueno:2003ei,Gil-Botella:2003sz,Gil-Botella:2004bv}
explores in some detail these possibilities: the main
conclusions~\cite{Gil-Botella:2004bv} are: if $\theta_{13}$ is large
enough ($>3
\times 10^{-4}$), one may distiguish inverted from normal hierarchy;
for $\theta_{13}<2\times 10^{-6}$, liquid argon may set an upper limit
on $\theta_{13}$; and for $\theta_{13}$ between these extremes, it may
be possible to measure it with liquid argon, even in a several kton
scale detector.  However, the supernova parameters are important to
the sensitivity to mixing: if the spectra of $\nu_e$, $\bar{\nu}_e$ and
$\nu_{\mu,\tau}$ neutrinos do not differ sufficiently, then one may
not be able to measure $\theta_{13}$ or distinguish normal from
inverted hierarchy, and larger scale ($\sim$100~ton) detectors may be
needed.

However, in combination with other detectors
(which may constrain both oscillation parameters
and the supernova physics) and perhaps with partial
knowledge of some of the parameters from e.g. long baseline
experiments, a liquid argon detector, with its excellent multi-flavor
capability, may yield significant constraints.

\subsection{Backgrounds}

The expected instantenous event rates from a nearby Supernova can be high  nevertheless the expected
backgrounds must be considered very carefully. The $200~kHz$ of cosmic muons is not a major
problem per se: a small tube surrounding every incoming cosmic ray particle (or shower initiated
buy it) can be easily removed from the fiducial  with a minimal loss of the signal rate thanks
to very high spatial granularity of the detector.

The dominant potential problem may be caused by the muon spallation and muon capture products,
an example being $^{40}Cl$ with a lifetime of 100 sec. All the muons stopping inside the detector 
volume are detected and the volumes containing potential capture products can be excluded from
the analysis. The size of the excluded region must include the diffusion and the accuracy with 
which a position of the end-point is known as a function of time, as a result of the fluid 
movement. The latter effect dominates as the velocity of the liquid circulation  reaches
up to $v=7~cm/sec$ thus giving up to $10~m$ displacements. The liquid circulation is laminar and
it is, in principle, well known. In practice a dedicated measurement system will be necessary
to validate the predictions and to keep the loss of the useful fiducial volume to the acceptable 
level.

A careful analysis of cosmogenically produced backgrounds as well as the backgrounds produced by 
incoming neutrons (the latter confined to the upper parts of the detector) is 
necessary to evaluate the real potential of the surface detector.

\section{Proton decay}

The question of nucleon stability is of fundamental importance and it has received a great deal
of attention the past few decades with large water Cherenkovs  in the forefront, but even 
a relatively small fine grained calorimeter, Soudan 2, making a significant contribution.

A 50 kton liquid argon detector would represent about two orders of magnitude step over Soudan 2,
at the same time providing more detailed information about the detail of a possible nucleon candidates.

\subsection{Motivation and Signatures}

A 50 kton water Cherenkov detector of  the SuperK is running since several years and its
limits on the proton lifetime keep improving with time. Further major improvements requires
a significant increase of the detector mass, perhaps to the level of Mton or so. This is
true for all of the decay modes where the efficiency of the water Cherenkov is very high
and this includes decay modes involving electrons and muons.

A niche for other detector technologies exists if proton decays primarily into K mesons.
Kaons at these energies are below the Cherenkov threshold in water, hence the the SuperK
detection efficiency is relatively low and the backgrounds are significant , as only secondary 
or tertiary particles are detectable.

\begin{figure}[t!]
\begin{center}
\includegraphics[scale=0.6]{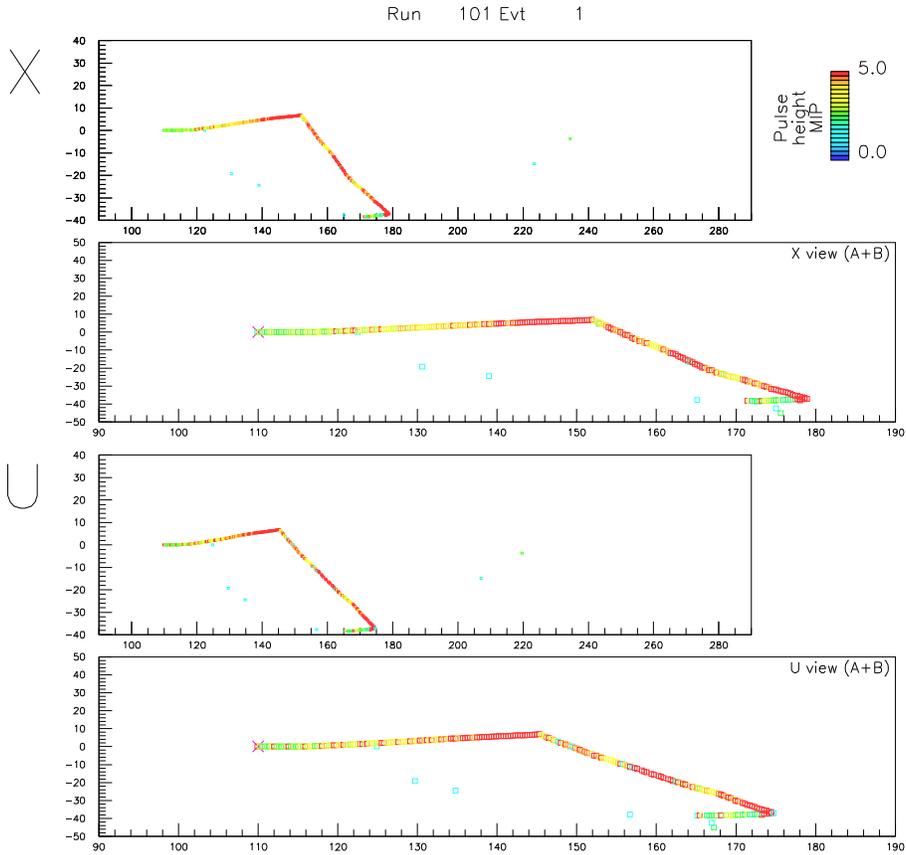}      
\vspace{0.5cm}
\caption[]{$0.5~GeV$ $K^{+}$ as detected in th liquid argon TPC. The color scale is adjusted such, 
that red corresponds to the ionization density  equivalent to 5 minimum ionizing particle.  }          

\label{ptoK}
\end{center}
\end{figure}

A $p\rightarrow K^{+}+\nu$ decay has been advertised  for a long time as primary example of
the analyzing  power of an imaging detector like a liquid argon TPC. Fig.\ref{ptoK} shows 
an example of a $0.5~GeV$ charged kaon as detected in the detector. The identification is
unambiguous and heavily over-constrained: kaon is identified by its decay chain $K^{+}\rightarrow
\mu^{+} \rightarrow e^{+}$. Kaon track is characterized by large ionization density, increasing
towards the end point. Muon track has well defined, and well measured but the range, momentum and
it has corresponding ionization density. An electron from the muon decay is easily identified and
well measured. The entire event occupies a relatively small volume of the detector, approximately 
a cube $0.5\times 0.5 \times 0.5~m^{3}$.   Not all instances of the kaon decays are equally spectacular,
some kaons decay at rest, some interact before the decay. Nevertheless, as the ICARUS studies have 
shown the efficiency for the detection and identification of the $p\rightarrow K^{+}+\nu$ decay
in the liquid argon detector exceeds $90\%$.

\subsection{Backgrounds}

It is widely assumed that the nucleon decay search requires detectors to be placed in deep
underground laboratories. There are two reasons for that:
\begin{itemize}
\item to reduce the rate of particles, mostly muons, crossing a very large volume of the 
detector to the
levels allowing reliable particle detection and reconstruction.
\item to reduce the rate of cosmic rays induced backgrounds, especially those induced by neutrons
\end{itemize}

The first concern is clearly dependent on the detector technology. A detector with good 
three-dimensional granularity, with no space charge effects can easily be operated on the surface and evidenced by the T300 module operation in Pavia. The nuisance of dealing with huge data rates
it he primary cost.

Far more important concerns is that of cosmics-induced backgrounds. The majority of cosmic rays
are muons. They, or any spallation products induced by them, do not constitute a significant
background sources as the energies expected for the proton decay products are very large, in hundreds
of MeV range. 

A serious concern is a neutron-induced  background. After all, neutron rates is only a factor 100
smaller than the muon rate. It is very unlikely that a search for $p\rightarrow e^{+} + \pizero$
can be conducted on the surface, for example. The decay modes involving kaons are different: 
the signal consists of a single charged kaon in the middle of a large detector. Strangeness 
conservation in strong interactions guarantees absence of such a background induced by neutrons.
To maintain good sensitivity to proton decay it is important that the background levels are kept 
below one event per year or so. The required rejection rate is huge and most unlikely processes
can easily swamp the measurement. An example of a possible background could be a neutron-induced 
associated production of $\Lambda + K$ with $\Lambda$ escaping a detection. It has been demonstrated
that this background does not require a very deep laboratory and the shallow depth of WIPP is adequate
to provide a background-free measurement\cite{cline} even without invoking any sophisticated   
analysis. An analysis of the surface detector requires far more work, and the ultimate result 
depends on the quantitative estimate of the probability of the produced $\Lambda$ escaping detection.
Given a very low detection threshold in the liquid argon the expected rejection factor will be
very large, but it is not known at the present time.

Even if the neutron-induced background is found to be negligible one needs to worry about $K^0$
component of the cosmic rays, as they can produce a single charged kaon in a charge-exchange reaction.

A significant background rejection factor against the background induced by the hadronic component
of the cosmic rays will be offered by the size and the analyzing power of the detector. It is 
very likely that the majority of neutrons and $K^0$ are produced in showering events and the 
accompanying  particles will be detected in the detector. A flux of 'isolated' hadrons is 
probably much smaller. 

The last line of defense could be a use of the top layer of the detector as a hadron absorber. 
It would  reduce the effective fiducial volume, but it could still offer a significant improvement
of the available limits or provide a discovery opportunity.  It is important to point out that
this upper region of the detector will provide a very good measurement of the actual backgrounds,
hence a defense of a possible positive result of the search would be relatively easy.

Given the magnitude of the necessary rejection factor against the cosmic rays background
a very detailed analysis of the performance of the proposed detector on one hand and the available
data on the hadronic component of the cosmic rays on the other hand is necessary to evaluate
reliably the physics potential of the surface detector.  
           
\chapter{Small Liquid Argon Detectors in the Intense Neutrino Beams}

Fermilab is blessed with two intense neutrino beams, the second one nearing its completion.
Unprecedented beam intensities offer very high interaction rates even with modest size
detectors, at the scale of 40 tons. Liquid argon TPC is a particularly suitable 
detector for the experiments in the near halls of MiniBOONE and NuMI beams. 
Simple topology/low multiplicity of the neutrino interactions  at these energies 
in conjunctions with good particle identification and good energy resolution will permit
complete kinematical reconstruction of the final states. Very low threshold of particles 
detection, few MeV for hadrons and sub-MeV for electrons, may be the deciding factor for 
several physics measurements. Relatively high density of liquid argon offers a good stopping 
power hence maximizing the size of the useful fiducial volume. 

Small detectors should be relatively simple to construct and operate as a scaled down
versions of the ICARUS modules, whereas their size will eliminate or greatly reduce
logistics and safety concerns. 

\section{Beam Spectrum and Beam Composition}

NuMI neutrino oscillation experiments aim at very precise determination of the
oscillation parameters. In a longer run they will become systematically limited 
by the understanding of the neutrino beam spectrum and its composition. Understanding
 of the $\nue$ component of the beam is  important for the appearance experiments.

The un-oscillated beam and its composition can be measured at a near detector position.
A reliable $\nue$ measurement is  difficult there, as the background from
$\pizero$-containing NC and $\numu$ CC events is very high and very different from the far 
detector. High rejection power of NC vs CC events and electrons vs $\pizero$'s, related to
very high spatial granularity of the detector, makes the liquid argon TPC especially
well suited for this purpose. 

Neutrino beam spectrum reconstruction requires good energy resolution and symmetric
resolution functions. Absence of long tails of the resolution function is especially
important in the regions where the energy spectrum is changing very rapidly. At low 
energies good neutrino energy calibration and resolution is limited by kinematical 
effects of rest masses and by the difference in response for different particles species.
Low detection threshold and good particle ID minimize these problems in case of a liquid argon
detector.    

\section{Neutrino Scattering Physics at Low Energies}
Experimental neutrino physics was in its infancy when low energy beams
were operational: beam intensities were very, very low, fluxes were
known very poorly, detectors were very crude or, in case of bubble
chambers, small and labor intensive. The world sample of events are in
the hundreds for some exclusive channels and in the tens of events for other
channels.  As a result, our knowledge of  neutrino interactions, even
the total cross section, in this energy regime is particularly poor.

The relatively low energies of current accelerator neutrino beams
(.1-20 GeV range) span the region below which the single pion
production cross section turns on, and above this, where deep
inelastic scattering (DIS) turns on.  With two neutrino beamlines at
Fermilab which span this rich energy range, and precision neutrino
detectors, a host of interesting physics can be addressed.

%

The difference in the neutrino fluxes on these two beamlines
translates to the breadth of physics that can be covered with programs
on both beamlines.  The Booster neutrino beamline's low energy flux
accesses low $Q^2$ interactions with a clean beam, free from DIS and
neutron backgrounds from high energy neutrino interactions around the
detectors.  The NuMI beam flux spans the interesting resonance to DIS
region with neutrino interactions up to 20 GeV.  Some of the wealth of
physics topics using these beams are described below.

\subsection{Neutrino Physics at NuMI Energies}
Neutrino Scattering in the NuMI beam spans the interesting transition
region from single pion production to deep inelastic scattering.
Onset of strange and charm particles production is taking place there
too.  This is the place where limits of applicability of perturbative
QCD may be studied.

Large samples of several millions of well reconstructed neutrino events, with complete 
kinematical reconstruction of well identified particles will be a veritable mine of bread 
and butter physics, especially because the forthcoming measurements of particle production (MIPP)
will enable neutrino flux prediction at the level of few percent.

A typical laundry list of physics topics includes, but is not limited to:
\begin{enumerate}
\item{Total $\numu$ and $\numubar$ Cross-sections -vs- $E_\nu$ notably below 
20 GeV}

\item{$\numu$ and $\numubar$ Quasi-elastic cross-section}

\item{$\numu$ and $\numubar$ resonance processes}

\item{Differential cross-sections, structure functions, 
and studies of PQCD and Non-PQCD; notably 
what is DIS and how does it link with non-scaling processes, such as 
QE and resonance}

\item{Cross-section of exclusive processes such as charm 
and strange particle production}

\end{enumerate}

Nuclear effects in neutrino interaction can be 
measured by using sheets of targets -from carbon to 
iron to lead - in the upstream end of the experiment liquid argon detector.

While these physics topics are interesting in their own right they are also of
great importance for the precise determination of of the oscillation parameters.

\subsection{Neutrino Physics at Booster Energies} 

The Booster neutrino beam provides high intensity neutrino beam with
energies below DIS turn-on.  A liquid argon time projection chamber
used to detect neutrino interactions in this beam can offer precision
measurements of cross sections as a function of $Q^2$, especially in
low $Q^2$ regime, thus allowing for determination of form factors and
study of nuclear effects.  

High intensity beams allow for study of low rate cross sections, such
as $\nu_e$-$e$elastic scatters, rich in interesting physics.  A very
important property of the LArTPC is a very low energy threshold for
electrons.  With low noise electronics and using the beam gate timing
it will be possible to detect electrons with energies as low as
$105~keV$. In conjunction with low energy and high rate of the
MiniBOONE beam this make such a detector particularly well suited for
studies $\nu_e$-$e$ elastic scattering and a measurement of the
neutrino magnetic moment.  We estimate that an experiment using a 100
ton detector can improve the existing limits by more than an order of
magnitude.

Listed below are examples of the physics topics available for study on
this beamline.

\begin{enumerate}
\item {Low $Q^2$ form factor measurements, in particular, a 
    measurement of the axial form factor and extraction of the strange
    part of the axial form factor}
\item{$\numu$ and$\numubar$ Precision measurement of resonant and coherent single pion channels}

\item{Study of nuclear effects in $numu$ and $\numubar$ cross-sections at low $Q^2$}
\item{Search for non-zero neutrino magnetic moment using $\nu_e$-$e$ elastic scattering}

\end{enumerate}

\section{Weak Mixing Angle}

Measurements of the weak mixing angle in the neutrino sector were among the
first experimental indications of the correctness of the electroweak model
which later advanced to the rank of the Standard Model. The same kind of measurements
started to cast some shadow on the this model or perhaps herald an arrival
of The New Physics\cite{nutev-weak}. Liquid argon TPC is particularly well
suited for studies of the neutral currents owing to the very low detection 
threshold and excellent muon identification. Very intense neutrino beams 
and good electron-$\pizero$ separation offers an interesting opportunity 
of the study of purely leptonic reaction: $\numu -e$ scattering. It is not
quite clear if the systematics can be controlled to the level required to measure
the weak mixing angle with the precision comparable to NuTeV, but a low
energy measurement may be interesting anyway. 

\section{Testing Ground For Neutrino Event Generators}

Neutrino oscillation experiments make heavy use of Monte Carlo simulations to extract
precise values of the oscillation parameters. Reliability of the extracted parameters
starts to be limited by the reliability of the underlying event generators which,
in turn, is limited by the scarcity of the experimental data. 

Very large sample of well reconstructed low energy neutrino events will provide
a stringent test for the neutrino event generators and will constrain various
ad-hoc methods of linking exclusive channels with deep inelastic regime.

Complete solid angle coverage and particle identification and good energy resolution
of the liquid argon detector will be of great asset here. 

\chapter{Liquid Argon Imaging Technology}

\section{Brief History}
The Liquid Argon Time Projection Chamber (LAr TPC), like many successful technologies, has 
several pioneers. Louis Alvarez  experimented with a liquid argon incarnation of 
a bubble chamber\cite{alvarez} as early as the late sixties. Emerging technology of liquid 
argon based calorimetry has stimulated studies of production and collection of 
electrons in liquefied noble gases and in particular studies of drifting electrons
over long distances in a liquid argon.  Electron drift over several centimeters was demonstrated 
 and  the 
relation between electronegative impurities (oxygen) and the drift distance was quantitatively 
established  in 1976\cite{hofman}.

First application of the liquid argon tracking detector to neutrino
experiments was proposed in 1976 by Herb Chen ~\cite{P496} as a
Fermilab proposal (P-496). This group demonstrated experimentally the
drifting of electrons in liquid argon over distances of 30
cm~\cite{chen_results}.  The subsequent R\&D effort at Irvine, Caltech
and Fermilab ~\cite{kephart} has lead to a program of developing
working prototypes of these Time Projection Chambers.

These liquid argon imaging detectors require low noise front end
electronics. Veljko Radeka, following a suggestion from Bill Willis, started
working in this area in 1975.  Consequently, the basic concepts of
electrode geometry and signal processing for a time projection liquid
argon ionization chamber were developed~\cite{radeka}.
     
A decisive step in the development of the liquid argon imaging
technology was a proposal by Carlo Rubbia~\cite{Rubbia} leading to the
formation of the ICARUS project.  Studies of electrons drifting over
large distances in solid and liquid
argon~\cite{drifting_solid,drifting_liquid}, construction and operation
of a collection of time projection chambers with increasing
mass~\cite{electron_image,ICARUS_RD,three_ton,three_ton_performance,140cm,Arneodo:vh}
and experimental studies of scintillation and Cherenkov light emission
in liquid argon~\cite{scintillation,cerenkov_light} were completed as
part of the R\&D efforts for ICARUS.  This work is complemented by
systematic studies of factors affecting the electron production and
lifetime in Argon~\cite{electron_lifetime,recombination} as well as by
a design of front-end electronics and data analysis
techniques~\cite{deltas,hit_finding,neural_network} again, developed by
ICARUS.  A major step was the application of industrial filters used
in the purification process~~\cite{filters} along with the
demonstration of feasibility of liquid phase
purification~\cite{liquid_purification}.  This progress has led to
electron lifetimes of several milliseconds, even in very large
detectors~\cite{purity_t600}.

The Liquid argon Time Projection Chamber is a well established and
mature technology.  The ICARUS experiment is the first example of its
application~\cite{imaging_device}. A successful design, construction
and operation of a large 600 ton module~\cite{T600} and the initial
results of the collected data~\cite{Amoruso:2003sw,Arneodo:rr} is a
convincing demonstration of the maturity of this technology.

Another contribution to the liquid argon imaging detector technology
has been a construction and successful operation of The Big liquid ARgon
Spectrometer, BARS~\cite{BARS}. This detector constructed for the
tagged neutrino beam at Serpukhov is a testimony to the fact that
detectors are usually easier to construct than neutrino beams.

\section{Principle of Detection Technique}

The following description of the detection technology and its practical
implementation is based on the ICARUS experience with the T600 module
in Pavia. It is only a review of a small subset of the available
information, judged to be the most relevant to the experiments
discussed in this letter.

\begin{figure}[t!]
\begin{center}
\includegraphics[scale=1.0]{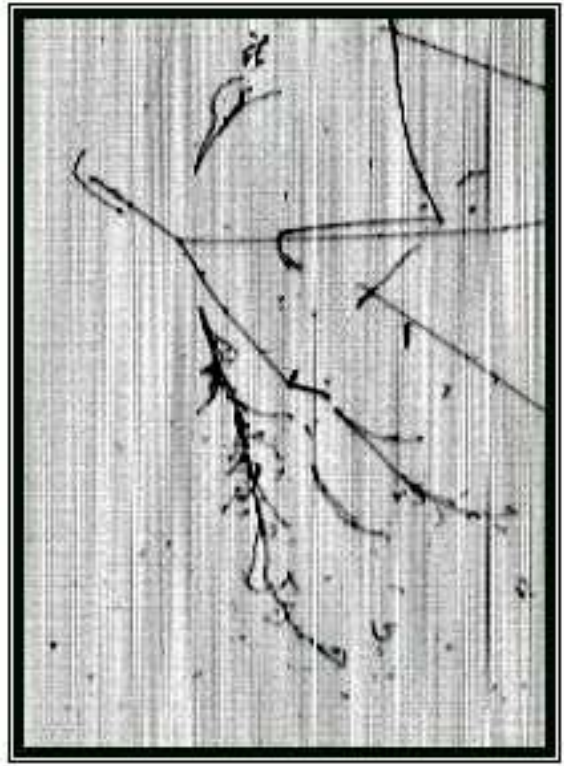}
\vspace{0.5cm}
\caption[]{Hadronic event registered in the T600 module of ICARUS. The detector
was operated in Pavia.}

\label{icarus_event}
\end{center}
\end{figure}

The analyzing power of the detector is  well illustrated in Fig.\ref{icarus_event}.
Its salient features are:
\begin{itemize}
\item All charged particles produced are detected. The detection threshold
  is 0.2 Mev kinetic energy (for electrons). Two track resolution is
  on the order of a few $mm$.
\item The three-dimensional view of the event is recorded with a voxel size on the 
order $3\times3\times0.5$ $mm^3$.
\item Ionization, dE/dx, is measured locally with an accuracy on the
  order of 10\%.  It allows for good particle identification.
\item Photon conversions are easily detectable with measurable
  energies. Radiation length in liquid argon is 14 cm (18 cm for
  photons)
\item Energy resolution for stopping particles is very good.  Energy
  resolution for electromagnetic showers is on the order of $\Delta
  E/E\sim 0.01/\sqrt(E)$; for hadronic showers it is on the order of
  $\Delta E/E\sim0.2/\sqrt(E)$
\end{itemize}

\begin{figure}[t!]
\begin{center}
\includegraphics[width=3.5in]{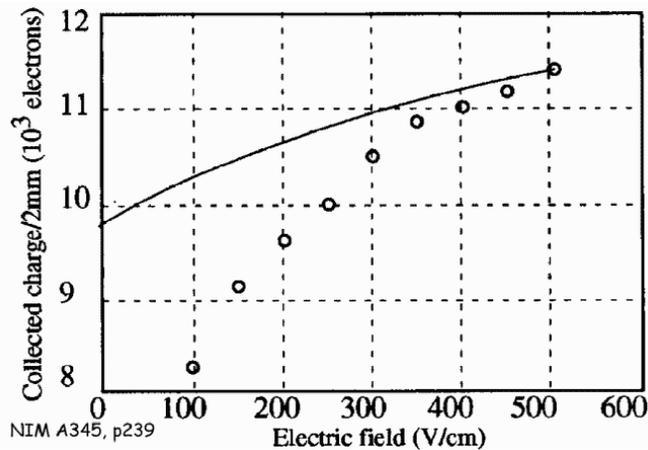}
\caption[]{Number of electrons collected per 2 mm of a minimum ionizing track~\cite{three_ton_performance}. 
Open circles represent data, solid line represents the variation expected from
a simple Onsager recombination model. }
\label{charge}
\end{center}
\end{figure}

\begin{figure}[t!]
\begin{center}
\includegraphics[width=3.5in]{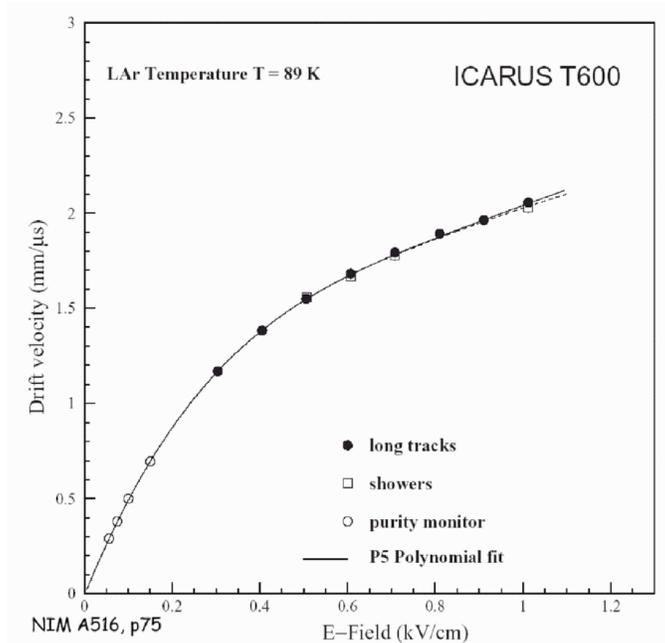}
\caption[]{Electron drift velocity as a function of the electric field in the drift volume~\cite{purity_t600}.}

\label{drift_velocity}
\end{center}
\end{figure}

A charged particle traversing the liquid argon produces electron-ion
pairs.  The number of ionization electrons produced per unit length of
the track depends on the operating voltage.  For typical operating
conditions at 500 V/cm, the number produce, $dQ/dx \sim 55000 e/cm$
see Fig.\ref{charge}. A uniform electric field throughout the detector
volume guides the electrons to collection electrodes. The drift
velocity of these electrons depends on the electric field strength,
Fig.\ref{drift_velocity} typically $1.5~mm/ \mu sec$. This field
strength leads to relatively long drift times up to 2 msec for 3 m
drift distances.  This, in turn, will set the requirements on the
purity of the argon.

From the measured drift time, the particle distance from the
collection wire can be determined, thus giving a two-dimensional
projection of the event, as shown in Fig.\ref{icarus_event}.  In
gaseous TPC's the position of the particle along the wire direction is
provided by the signal induced by the avalanche on the the set of pads
positioned behind the collection wire. Absence of an avalanche in
liquid argon makes this technique unworkable here. Instead, a wire
grid (or grids), placed in front of the collection wires, readout the
signal induced by the passing electrons.  With proper biasing these
grids can be made transparent to the drifting electrons and thus not
attenuate the signal to be recorded on the collection wire\cite{grids}.
This technique permits more than one additional coordinate to be
obtained and provides redundancy for three dimensional reconstruction.

\section{ICARUS T600 Module}
The ICARUS T600 module is a dual 300 ton detector constructed as
proof-of-principle, demonstrating the viability of a large scale
liquid argon detector.  Its design is tailored to the requirements of
the underground experiment in the Gran Sasso laboratory. In
particular, the rectangular module, designed with thin thermal
insulation, can be constructed elsewhere, transported over the public
highways to the laboratory, and is sized to fit through the
laboratory's entrance gate.

The successful construction and operation of the T600 module is the
primary evidence of the maturity of the technology.  A large fraction
of the technical solutions learned in building the T600 are directly
applicable to detectors of different scale, both larger and smaller.
However, some aspects, particularly those related to the cryostat
construction and the cryogenic systems, are specific to the Gran Sasso
experiment. We will mention them very briefly here.
 
\subsection{Cryostat and Cryogenics}

The design of the cryostat and the cryogenic system for the T600 is
driven by the requirement that the LAr container have to be
transported through the Italian highways and to pass through the entrance
gate of the underground Gran Sasso Laboratory~\cite{T600}.

Aluminum containers $4.2 \times 3.9 \times 19.9$ $m^3$ are built out
of perforated aluminum honeycomb panels with embedded liquid nitrogen
cooling pipes, sandwiched between aluminum skins.  Thermal insulation
is achieved through two layers of $20 cm$ thick Nomex honeycomb
structures on both sides of a perforated aluminum cold shield.

The cooling system is designed to keep the liquid argon temperature
stable to $1^o$ C over the entire detector volume. It is based on 
circulation of liquid nitrogen pressurized to 2.7 bar with a rate
of 8 ton/hour/container.

\subsection{Argon Purification}

High argon purity is a fundamental pre-condition of the functional
detector.  For drift distances on the order of $2-3$ meters it is
necessary to maintain an electron lifetime of several milliseconds in
argon and this, in turn, requires that contamination of
electronegative impurities be maintained at the level of $10^{-10}$
$O_{2}$ equivalent.

Purification of liquid argon from an initial purity level of about
1~ppm to the level of $0.1~ppb$ can be achieved by recirculation of the
argon through a combination of commercial filters: Hydrosorb and
Oxisorb~\cite{filters}.

The T600 module is instrumented with two gas recirculation systems
with nominal capacity of $25$ $m^3/hours/system$ and one liquid phase
recirculation unit with capacity of $2.5$ $m^2/hour$.

The purification filters are designed for gas phase purification.
However, dedicated studies~\cite{liquid_purification} have
demonstrated that they perform well for liquid argon as well. In fact,
liquid phase purification is much more efficient due to the fact that
a larger mass of argon can be purified in the same amount of time. It
has also been demonstrated that the purity level of the liquid is much
higher -- by a factor of 100 or more -- that the original gas as a
result of impurities freezing out at the cold surfaces of the liquid
container.

The purity level of liquid argon is a result of interplay of several
effects:
\begin{itemize}
\item removal of impurities by the purification system
\item inflow of oxygen due to potential leaks
\item out-gassing from warm surfaces in the gas phase of the container, primarily the cables
\end{itemize} 

Analysis of the performance of the purification system
~\cite{liquid_purification,purity_t600} indicates that the contribution
from leaks to the impurity level was negligible while the purity level
improving steadily as a result of the purification process (liquid
phase mostly).  Although the purity level achieved during the test run
corresponded to an electron lifetime of about $2$ $msec$ the observed
purification rate indicates that the ultimate purity level would
correspond to a lifetime of $13~msec$~\cite{purity_t600}.

\subsection{High Voltage}

The high voltage system must produce a uniform electric field in the
drift volume. The T600 configuration contains a cathode plane in the
middle of the container with collection wires on each side of the
container, thus giving a maximal drift distance of $1.5~m$. The module
is operated with a drift field of $500~V/cm$, hence the central
cathode runs at a voltage of $75~kV$. With future applications in
mind, a short successful test run was performed with the high voltage
operating at $150~kV$.
 
High voltage for the T600 is supplied by a commercial HV power supply
made by Heinzinger.  A novel design custom-made HV feed-through was
designed and built for the specific geometry of the T600 module.  This
design can be easily applied to other cryostats too.

The uniformity of the electric field is assured by a field cage
consisting of a set of rectangular frames constructed from stainless
tubing and spaced by $5~cm$ along the drift direction. . The frames
are biased to potential linearly decreasing with a distance from the
central cathode with a resistor chain. Four independent resistors,
$100~M\Omega$ each, are employed at every step for redundancy. The
resulting current drawn from the HV power supply is $1~mA$.

\subsection{Inner Detector}

Electrons produced by charged tracks in the detector volume drift to
the sides of the module volume and are detected by 3 planes of wire
chambers.  The first two planes -- so called induction 1 and induction
2 -- register signals induced by the electron cloud traversing the
wire plane.  The third plane -- the collection plane -- collects all
of the electrons drifted to it. The position of the wires registering
a signal in conjunction with the measured drift time gives two
coordinates of the originating track. This along with the information
from the two induction planes, oriented at stereo angles $+60^o$ and
$-60^o$ with respect to the collection wires, thus provides three
dimensional information.  All three position measurements from a given
track segment have the same timing information, thus facilitating the
three dimensional track reconstruction.  Wire spacing within each
plane as well as the distance between the wire planes is $3~mm$.  The
planes are biased by about $300~V$ with respect to neighboring planes.

The chief requirement of the design of wire chambers is to assure
survivability of all the wires during the thermal cycle of the
detector cool-down.  The problem is somewhat complicated by a
difference of the thermal expansion coefficients of different
structural materials used in the construction (steel and aluminum).
The chambers are constructed using $150~\mu$ thick steel wires
tensioned with $5~N$ (horizontal wires) or $12~N$ (stereo wires).
Chambers are constructed in a modular fashion with wires cut, spooled
and cleaned at remote laboratories, and later unwound and mounted
inside the detector.

A very elaborate tensioning system sets the initial tension of the
wires and compensates the tension increase due to thermal cycle.  A
thermal cycle of the entire system was tested using a smaller, 10 ton,
ICARUS prototype.  None of the 1900 wires broke or even reached their
elastic limit~\cite{Arneodo:vh}.
 
Twisted-pair flat cables are used to carry the wire signals out of
the cryostat to the front-end electronics located on top of the
detector. All signal feed-throughs are therefore at room temperature.

\subsection{Readout Electronics}

Liquid argon time projection chamber readout consists of a digitized
waveform for every wire.  Given a drift velocity of $v_{drift}=1.5~mm
/\mu sec$, a sampling rate on the order of $2~MHz$ is sufficient for
waveform digitization. Because the signal size small, of the order of
15,000 electrons, the most critical parameter of the readout
electronics is input noise.

The analog stage of the electronics is a commercial 32 channel board
CAEN V791. It consists of a 32 preamplifiers based on the analog
BiCMOS chip, a 16:1 multiplexer and a 40 MHz 10 bit FADC.  There are
two versions of the V791 boards with different shaping times for the
collection/first induction plane and for the second induction plane.
The latter conditions a bipolar induced signal into a signal with very
similar characteristics to the signals from two other planes.

The electronics noise is dominated by the input stage of the first
amplifier and it is proportional (with an offset of 500 electrons) to
the capacitance of the detector. For a typical channel with a
capacitance of $400~pF$ the signal-to-noise ratio exceeds a factor of
five for a minimum ionizing track.

Full digitized waveforms for all wires make for a very large data
volume.  This data set can be reduced in the digital stage of the front-end
electronics, V789 board, by a dedicated ASIC performing hit finding.
The resulting data (reduced or not) are stored in 128kB VDRAM and
eventually, if triggered, transferred to the data acquisition
system.

\subsection{T600 Operation in Pavia}

The complete T600 detector module was commissioned and successfully
operated for several months in Pavia. The main focus of this technical  run was a
checkout of the entire system including the cool-down of the cryostat,
purification of the liquid argon, and check of the detector
performance.
 
Over the course of the run, tens of thousands of selected cosmic ray
events were recorded including those with long horizontal tracks~\cite{Arneodo:rr} or high
energy cascades.  The collected data has provided yet another
demonstration of the imaging capability of the liquid argon TPC. It
has also demonstrated that the cosmic ray and environmental background
in the surface detector can be separated from other interactions in the volume.

The analysis of the data provides a valuable feedback for the software
development, and has produced the first physics
results with this technology~\cite{Amoruso:2003sw}.

\chapter{Liquid Argon Off-axis Detector}

The ICARUS collaboration has demonstrated fundamentals of the liquid
argon imaging technology:
\begin{itemize}
\item large quantities of liquid argon can be purified using commercial
filters to the levels permitting drift of electrons over several meters
\item large area wire chambers can be constructed in a manner ensuring 
their survivability of the cool down process and their operation at
liquid argon temperatures 
\item low noise electronics can be designed and constructed providing
a signal-to-noise ratio for detector capacitances up to $400~pF$

\end{itemize}

The ICARUS collaboration  has also demonstrated that a large scale
implementation of such a technology is possible even with demanding
constraints of the underground laboratory. It is a testimony to the
success of the ICARUS project that attempts to design detectors at much
larger scale are made quite frequently in the framework of neutrino
factories\cite{large_detectors} and/or future large underground 
laboratories\cite{LANNDD}.

\section{Overview}
With the fundamental principles proven to be well understood the principal challenge
of a design and construction of a very large detector is of the engineering
nature and it reduces to a problem  of a design of a safe, cost effective and practically
realizable detector. It should be pointed out that a possible scheme could involve
construction of a number of ICARUS-like modules. This is not a cost effective solution, 
though.

It has been recognized that the a large detector constructed inside a single cryogenic 
vessel represents much more attractive possibility\cite{large_detectors}.  In this chapter
we present a design of a possible detector using such a concept. This design utilizes 
maximally the solutions used in the construction of the ICARUS module. There are several
possible improvements, which can be incorporated in the design once the necessary R\&D
is successfully completed. We describe these possible developments in Chapter 8.

In the following we demonstrate that it is possible to construct  a cost-effective 
detector large detector using the existing technology. As an example we use take a 50 kton 
surface detector constructed in off-axis position in a NuMI neutrino beam. Many of the 
industrial aspects (tanks, cryogenics, liquid argon) were studied by G. Mullholand from
Applied Cryogenics Technology for the LANNDD project\cite{mulholland}.
  
\section{Liquid argon tank and the cryogenics}

Large cryogenic tanks are built by industry to store liquefied natural gas (LNG).
Several vendors offer a large variety of tanks with volumes up to $200,000~m^3$.
For example purposes in the following we use Chicago Bridge and Iron\cite{CBI}, but
several other vendors exist.

It is interesting to note that the thermodynamics of storage of liquid argon is 
quite similar to that of liquefied methane. The boiling points of argon and methane
are $87.3~K$ and $111.6~K$ respectively. The heat of vaporization per unit volume
of both liquids is the same within $\sim5\%$ therefore the volumetric boil-off rates
will be comparable for the same thermal insulation. The main difference between the storage
of liquid argon and that of liquid methane stems from the difference in their densities:
the liquid argon tank need to withstand 3.3 times higher hydrostatic pressure.

\begin{figure}[t!]
\begin{center}
\includegraphics[scale=0.6]{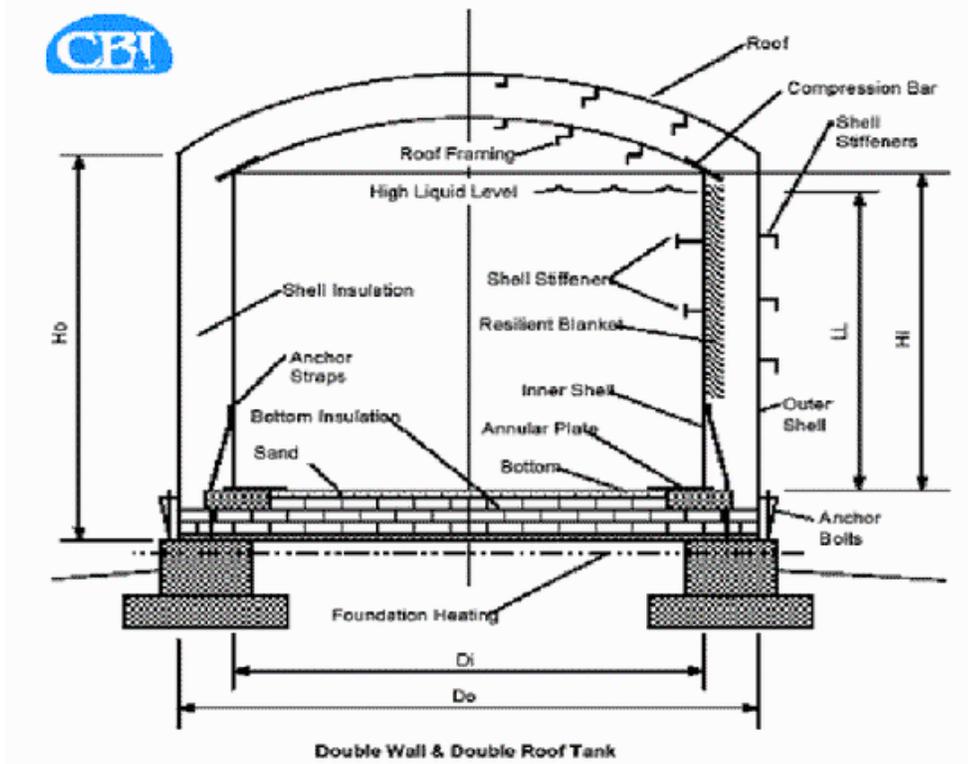}
\vspace{0.5cm}
\caption[]{Double wall containment cryogenic vessel, CBI standard design}

\label{CBI_tank}
\end{center}
\end{figure}

The tank necessary to store $50~kton$ of liquid argon has $35,000~m^3$ volume and it
would be a cylindrical tank $30~m$ high with $40~m$ diameter. There are several possible
designs of storage vessels\cite{CBI}. The one suitable for our application is that
of a double wall and double roof vessel illustrated in Fig.\ref{CBI_tank} 

To ensure reliability of the tank at the cryogenic temperatures the inner vessel will be constructed from
full penetration butt welded nickel steel plates (Fe9NI, A553, Type I). The outer vessel is constructed
out mild steel.
 Thermal insulation is provided by
$1.2~m$ thick layer of Perlite purged with dry nitrogen.
This thermal insulation implies that the boil-off rate of liquid argon will be in the
range of $0.05\%~/day$ or $25~ton/day$.

It is very important to stress that this kind of tank is not capable of taking any significant
external loads and in particular it cannot be evacuated. 

Such a tank is very similar to the tanks constructed for gas industry, hence
the construction costs can be reliably estimated. CBI quotes an estimated cost of turnkey system
(excluding external piping, pumps, refrigeration system, electrical,
 instrumentation and controls) is \$11 M. This cost does not include penetrations required for
High Voltage and signal feed-troughs but they are expected to be well within the accuracy of
the quoted price, which is about $20\%$.

\section{Liquid argon and the purification system}

\subsection{Liquid Argon}
Liquid argon is a by-product of air liquefaction. It is widely used in industry for heat
treatment, sintering or  as a shield gas. An annual production of the liquid argon in the US
is about $1,000~kton$. Specification of industrial grade liquid argon lists an oxygen content
of $2~ppm$, although in practice the purity levels routinely seems to be better than $1~ppm$\cite{Rahm}.
There exist grades of high purity, 0.1 ppm, of liquid argon but they do not present a cost effective
avenue for a large experiment, especially that it may be difficult to maintain such a high purity
level during the delivery and filling process. 

The typical price of the liquid argon is of the order of \$0.40/kg  but the delivery costs may
be quite significant. Following the study for the LANNDD\cite{mulholland} we  use \$0.60/kg as an estimated price
for delivered liquid argon. This cost assumes that the argon is delivered by trucks. Significant savings
can be realized if the experiment location allows for the delivery by railroad cars. In the following we will use
a figure of $\$30~M$ for the cost of $50~kton$ of argon.

\subsection{Cryogenics}

\begin{figure}[t!]
\begin{center}
\includegraphics[scale=0.6]{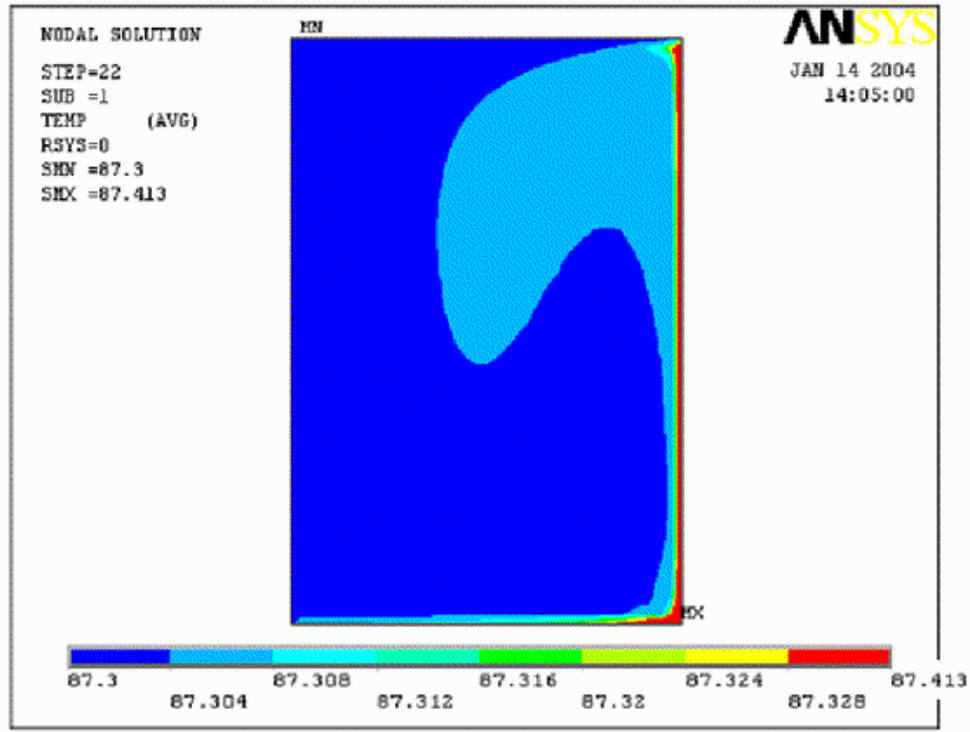}
\vspace{0.5cm}
\caption[]{Liquid argon temperature distribution in a axisymmetric tank. One half of the tank is shown.}

\label{temperature}
\end{center}
\end{figure}

Thermodynamics of a liquefied gas storage tank is quite different from that of a cryostat with active
cooling system. The storage tank is, to a good approximation, a big boiling pot. The tank walls are
the heat sources. Warm(er) argon rises along the walls to the top where it cools down by evaporation and
sinks to a bottom in the middle part of the tank. These convection currents stabilize the temperature
of the entire volume to be uniform within $0.01^o~K$ over the most of the volume\cite{tang_1}. The temperature along
the walls is $0.1^o~K$ higher than the boiling temperature along the walls of the tank. The distribution
of the liquid temperature throughout the tank volume is shown in Fig.\ref{temperature}.

The boil-off gas constitutes $0.05\%/day$ of the total mass of the argon, or $25~ton$ in our specific case.
The cryogenic system is necessary to re-liquefy this gas mass in order to avoid losses of the material.
The liquefied argon is pumped back into the storage tank. This system ensures that the temperature
of argon never falls below that of the freezing point, which is only $3^o~K$ below the boiling point.

To bracket the cost of the refrigeration unit we use a specific proposal of Cosmodyne Corp. for the
LANNDD detector. The estimated cost was $\$2.8~M$ for a unit with refrigeration capacity of $100~ton$
of liquid argon per day. 

\subsection{Purification}
Argon purity is of the paramount importance and it is critical that the adequate purification system
and procedures are in place to ensure that it is achieved.

The problem breaks out into several independent components:
\begin{itemize}
\item the delivered argon with purity levels at 1-2 ppm must be purified to below 1 ppb level
\item tank walls as well as all the materials placed inside must be cleaned up to high vacuum standards
\item all gas, especially oxygen, must be removed from the tank volume
\item the tank with all its penetrations and feed-troughs must maintain its hermeticity and present
no gas leaks.
\item out-gassing products must be removed from the tank volume faster then they are produced
\end{itemize}

To set the scale of the problem we point out that 0.1 ppb purity level of liquid argon
corresponds to $15~l$ of air inside the tank volume.

The inner tank surfaces, after installation of all structural elements, will be washed using
diluted solution of nitric acid, rinsed with de-mineralized water and dried with dry air.
The tank will become a clean room and the further operations will adhere to the clean room
standards. 

The tank of this construction does not permit its evacuation as a means to remove the oxygen.
Instead, several volume exchanges with dry nitrogen will reduce the oxygen level to a level of about $1\%$. 
The purification process will continue with several purges of the tank volume with gaseous argon
injected at the bottom of the tank. Due to  higher density of the argon the gas volume will be stratified
with argon gradually pushing the oxygen and nitrogen out of the tank volume. This process will 
reduce the oxygen level in the tank below $0.1\%$ level, or in absolute terms, the amount of 
oxygen in the tank will fall below $50~kg$.    
    
It is expected that the argon will be delivered to a central location from where it will be
pumped to one of the three intermediate storage tanks with 20 ton capacity through a set of
purifying Oxisorb/Hydrosorb purifying filters. These tanks will be equipped with argon 
purity/electron lifetime detectors. If necessary the liquid argon will be recirculated at the
rate of $3~ton/hour$ through the same set of filters until the desired purity level is achieved.
Batches with abnormally high impurities level, if any, will be rejected. Only  argon with the purity
level comparable to that in the main tank will be transferred to the main tank volume, higher at
the beginning of the fill and falling gradually to $0.1~ppb$ towards the end.

Pure argon transferred to the main tank volume will be contaminated by the residual oxygen in the
gas volume and it needs to be purified further and this will be provided by the dedicated purification
systems. At the initial phase of the fill the oxygen contamination will be quite high, thus 
a simple distillation column will provide a cost-effective means of purification. We envisage
a system of multiple heated columns with water-cooled heat exchanger with a flow capacity of 
$170~m^3/hour$, or $240~kg/hour$, which would provide a total gas volume exchange every 6.5 days at the initial phase,
faster at the later stages as the tank volume is filled with liquid. This system would help to
remove products of the out-gassing of the detector elements, mainly cables, which will be most 
abundant in the initial phase. It is likely that it will not need to be operated during the operation
of the detector. A cost estimate for the construction of such a purification system
(PRAXAIR) is $\$700 K$. 

To attain the desired purity level we envisage a liquid purification system using standard 
Hydrosorb/Oxisorb combination with a capacity of $100~ton/hour$. This system will permit a complete
liquid volume exchange in a few days in the initial phase of the fill procedure, when the total mass
of the liquid is relatively low, thus permitting efficient purification. In the later phases of the
the fill process the purity standard for the liquid transferred to the the tank volume will be
raised accordingly on one hand and the impurities level will be reduced by the increase of the
purifying effect of increased cold surfaces on the other hand. 

We expect that the filling process of the tank will take about 1 calendar year. This implies handling
of $200~ton$ of liquid argon per day. This is comparable with the capacity of liquid nitrogen
handling at Fermilab.

It is expected that during the operation phase of the experiment the primary function of the
purification system will be to maintain the purity level, especially by removing the out-gassing
products in the gas volume of the tank. For this purpose we envisage a gas purification system with
a flow capacity of $1000~m^3/hour$.

The design and sizing of the purification system is based on the operational experience of the ICARUS 
T600 module. We expect that the volume-to-surface ratio should make the purification much easier
in the large detector, hence probably some down-scaled system will be quite sufficient. On the other hand,
given the importance of the argon purity for the functioning of the detector, we feel that substantial
overcapacity is necessary to ensure reliable operation.

The cost estimate for   truck stations, three twenty-ton tanks, refrigeration for the three tanks,
purifiers for the twenty ton tanks, pumps to empty the tanks into the big
tank, vacuum insulated transfer line from the small tanks to the large tank,  pumps and
gas handling for a 100 ton/hour liquid purifier and a 1000 cubic meter/hour
purifier is $\$6.5~M$. The necessary Hydrosorb/Oxisorb filters will cost abut $\$2~M$.

The total cost of the handling/refrigeration and purification system is estimated to be $\$12~M.$

\section{Mechanical Support Structure}

\begin{figure}[t!]
\begin{center}
\includegraphics[scale=0.3]{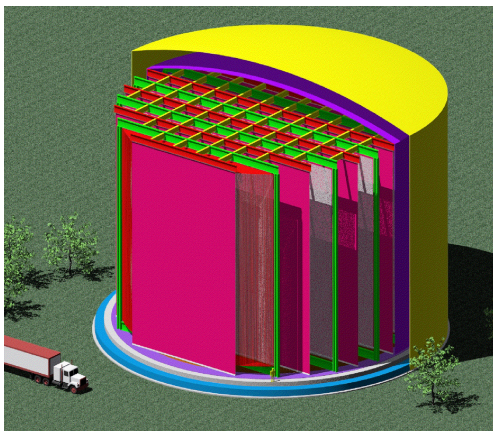}
\vspace{0.5cm}
\caption[]{Overview of the detector inside the tank. Green planes are the wire chambers, cathode planes are
in pink, field shaping frames are red (only one set is shown for clarity) . 
Most of the volume is free of instrumentation and it is filled with liquid argon. [Bartoszek Engineering]  }

\label{detector}
\end{center}
\end{figure}

The inner detector, wire chambers, cathode planes and field shaping electrodes will be supported from
above by a truss formed by a set of I beams as shown in Fig.\ref{detector}.

 The total load will be transferred to the floor of the
building by vertical beams at the periphery of the tank.

The truss will be use also to support a flat roof of the inner tank the insulation layer of Perlite
and a gas-proof membrane. The outer tank will have a domed roof as in the standard tank design.
The inner membrane will be locally reinforced to provide the support for front end electronics 
and the access path.

Such a flat inner roof/support structure will not be able to take a significant load owing to a very large
surface area of $1200~m^2$. A system of relief valves and a control of the argon liquefaction process
will ensure that the pressure differential of the domed volume and the gas volume of the inner tank are
within the allowed range. Even for small pressure differential, however, the displacement of the inner roof 
and the supporting I-beams will be significant. It is therefore important that the inner detector design
allows for vertical displacement of its support structure.

Given the size of the support structure, the requirements concerning materials purity and the cryogenic
operating conditions, detailed engineering studies are necessary for a reliable cost estimate. 
At the moment we have evaluated the projected cost using \$25.0/lb as an estimate of the construction cost, 
including materials and labor costs. The total cost is estimated to be \$37.7~M and Table \ref{support_cost} 
gives more detailed breakdown.  

\begin{table}
\begin{center}
\begin{tabular}{||l|r|r|r||}
\hline \hline
 Element  & Quantity  & Total weight, lbs &Estimated cost, K\$  \\ \hline \hline
Columns $W14\times 730$ + gusset & 12 & 1,234,248 & 30,856.20 \\ \hline
WC1 Top Beam - $W44 \times 224$ & 2 & 59,732 & 1,493.30 \\ \hline
WC2 Top Beam - $W44 \times 224$ & 2 & 53,798 & 1,344.95 \\ \hline
WC3 Top Beam - $W44 \times 224$ & 2 & 39,400 & 985.00 \\ \hline
Roof Truss short beams      & 96 & 119,616 & 2,990.40 \\ \hline
\hline
Total & & & 37,669.85 \\ \hline
\hline

\end{tabular}
\end{center}
\caption{Estimated cost of the elements of structural support of the inner detector, assuming \$25.00/lb for
the finished product.}
\label{support_cost}
\end{table}   

\section{High Voltage}

\begin{figure}[t!]
\begin{center}
\includegraphics[scale=0.3]{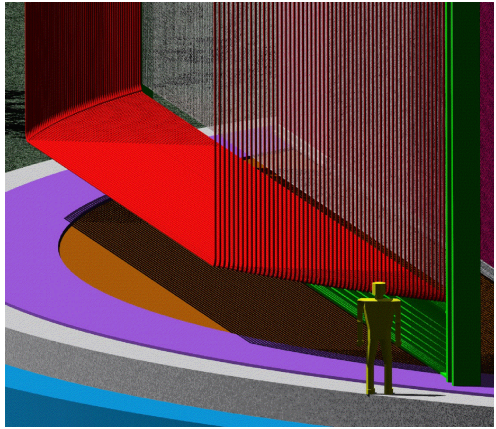}
\vspace{0.5cm}
\caption[]{The drift volume and the field shaping cage. The green beam supports the wire chamber, the cathode
plane at the left edge  is removed for clarity. [Bartoszek Engineering]  }

\label{field_cage}
\end{center}
\end{figure}

An uniform electric field of $500~V/cm$ in the drift volume will be provided by a set of 7 cathode
planes spaced $6~m$ apart and a collection of field shaping electrodes formed by stainless steel tube
frames positioned every 5 cm between the cathode and the wire chambers as shown in Fig.\ref{field_cage}.

Seven independent HV power supplies with $150~kV$ rating will be used. HV feed-through similar
to the ICARUS design will be used, the main difference being the length of the ultra-high molecular
weight polyethylene insulator. The HV feedthroughs will be positioned at the periphery of the tank,
nevertheless they must penetrate the outer roof, the thermal insulation layer and the gas volume in the
inner tank. It is likely that the total length of the necessary insulator will be of the order of $3 m$.

The cathode planes will be constructed from thin perforated stainless steel sheets and suspended from
the I-beams through a set insulators. The liquid argon level control will be constructed to ensure that
the cathodes are always fully submerged in the liquid.

Uniform electric field is achieved by linear gradation of the potential in the space between the cathode
and the wire planes. This is achieved with a set of frames constructed from the stainless steel tubes.
Four independent resistor chains degrade the potential in a linear fashion. The dimensions of the field 
shaping frames conform to the local tank dimensions. The frames are supported from the insulated beams
perpendicular to those supporting the cathodes and the wire chambers. The bottom of the frame is supported
off the floor through a set of insulating stands.

The total cost of the High Voltage system is dominated by the construction cost of the cathodes and the 
field shaping frames. Detailed engineering studies are necessary for a reliable estimate. Using, as before,
\$25.00/lb as an estimate of the total construction costs we arrive with the expected cost of \$5.7~M.
A more detailed breakdown is shown in Table \ref{cost_HV}.

\begin{table}
\begin{center}
\begin{tabular}{||l|r|r|r||}
\hline \hline
 Element  & Quantity  & Total weight, lbs &Estimated cost, K\$  \\ \hline \hline
Cathode top beam 1, $W36\times 135$ & 1 & 18,313 & 457.83  \\ \hline
Cathode top beam 2, $W36\times 135$ & 2 & 34,912 & 872.80  \\ \hline
Cathode top beam 3, $W36\times 135$ & 2 & 29,210 & 730.25  \\ \hline
Cathode top beam 4, $W36\times 135$ & 2 & 15,764 & 394.10  \\ \hline
Field shaping electrodes & 708 & 131,579 & 3,289.47  \\ \hline \hline
Total & & & 5,744.44 \\ \hline
\hline

\end{tabular}
\end{center}
\caption{Estimated cost of the elements of the HV field cage, assuming \$25.00/lb for
the finished product.}
\label{cost_HV}
\end{table} 
        
\section{Wire Planes}

Drifting electrons will be detected/collected with a set of wire planes positioned in the middle between
the cathode planes. The drift distance is $3~m$. Two back-to-back chambers are detecting
electrons coming from opposite directions. Every chamber is composed of three wire planes: $+30^o$, $-30^o$
induction planes and a collection plane with vertical wires. The collection plane is furthest away from the 
cathode plane.

The wire lanes have very large dimensions: the biggest one is $30\times 40~m$ with the wire lengths 
ranging up to $36~m$ and their construction will present a significant engineering and technical challenge.
The main issues are:
\begin{itemize}
\item stringing and tensioning of very long wires, 
\item large wire capacitance which determines the electronics noise, 
\item large forces on the support frame,
\item the survivability during the cool-down phase of the experiment
\end{itemize}
 The most critical issue is quality control of the produced chamber: wire replacement 
is not possible on one hand and a single broken wire can short the entire drift gap thus taking a 
significant section of the detector out of operation.

To minimize the electronics noise it is important to keep the detector capacitance at a minimum.
In ICARUS geometry, with $3~mm$ wire spacing and $3~mm$  spacing between plane the capacitance of the wire is
$20~pF/m$. Wire spacing determines the frequency of sampling of charged particle trajectory and the minimum 
track length detectable as a trajectory. For relatively simple topologies of neutrino interactions in
the energy regime of the NuMI neutrino beams a $5~mm$ wire spacing is fairly adequate. Increasing the
wire spacing to $5~mm$ increases the minimum ionizing particle signal size by $60\%$ and it decreases
the wire capacitance (and the electronics noise associated with it) by $30\%$\cite{tang_2}, in 
comparison with the ICARUS case. It is worth mentioning that
the wire capacitance varies approximately like $1/r$ and  further increase of the wire spacing offers
very small  reduction of the capacitance.

Since there is no gain at the wire large diameter wires can be used in the detector construction.
We plan to use $150~\mu$ thick stainless steel wire. To ensure the wires' stability they will be  
tensioned with 10 N force, similar to that of the ICARUS T600 module,  and constrained with
spacers  every $5~m$. As the result 
the chamber frame will undergo a compressive force of about $10,000~N/meter$, $\sim1~ton/m$, 
 or $40~tons$ for the longest 
chamber. This will require a supporting beam with adequate stiffness. Additional problems of wire
construction, ensuring the tension uniformity and survivability of wires during the cool-down phase 
are addressed by the  construction technique, described below.

The wires will be grouped in batches of 16 wires.
The 16 individual wires will be attached with a slipknot to stainless pins and inserted into a 
PEEK board (side A)
which will determines the wire positions. A required length of 16 wires will be wound on an auxiliary spool
with spacers inserted every $5~m$. The ends of the wires will be threaded through another PEEK bar (side B) 
which will ensure proper  wire spacing an the other end of the chamber and attached with a slipknot to 
a PEEK ring.

The spools will be positioned on the upper level of the supporting beams inside the detector tank and
unwound dropping the PEEK board (side B) to the floor where The PEEK board will be attached to a horizontal
bar determining the wire position. Below the positioning bar there will be sets of tensioning weights
$1.3~kg$ each attached to stainless string with hooks. Every wire of the chamber will have its own weight
attached to the PEEK ring at the end of the wire. This system will result in:
\begin{itemize}
\item a simple and practical construction technique. The required accuracy of the wire length at the
construction phase will be at the level of several $cm$, limited by the vertical space for movement
of the weights. This may be contrasted with the  precision of $100~\mu$ required for the ICARUS chambers. 
\item proper tension for every individual wire. This tension will be  constant during all phases of 
construction, independent of the
progressively larger deflection of the support beam as the as large portions of the wire plane are installed.
\item simple procedure to test by over-stressing of every wire   by putting an additional test weight 
on it for a short while

\item no over-tension and no risk of wire breaking during the cool-down process. Since the wires have a  
smaller heat capacity than the supporting structure they
will  shrink earlier that the frames but this will only result in a change of the heights of 
the weights.
\item the regularly spaced pins of side A allow for a simple mass terminated connection of the 
electronic readout cables

\end{itemize}

The procedure described above works well for the vertical plane of wires. For the stereo wires the 
modifications are minor: the bottom PEEK board (side B) must be displaced horizontally by $17.3~m$ before
attachment to the positioning bar. The PEEK board will have shaped channels to guide the wires 
through a $30^o$ turn before the attachment to the tensioning weight. 

The chamber construction technique described above will have to be appropriately modified for the
installation of stereo wires that either originate or terminate on the side support beams.

The cost of the chamber construction is likely to be dominated by the labor necessary to string approximately 
300,000 wires, and this should be largely a function of number of wires only since the length-dependent 
component of the 
labor is rather small.  

 At present the best estimate is that it will
cost about $\$0.8~M$ to construct a single chamber (6 planes), hence the
total cost of chamber construction is expected to be $\$5~M$. 

\section{Electronics}

ICARUS collaboration has developed a complete chain of readout electronics which was demonstrated
to perform adequately for a large surface detector. There are several decisions to be made in the
electronics area, the primary one being the placement of the electronics inside or outside the 
cryostat. These decisions have repercussions for many other detector elements like chamber geometry,
cables, feed-troughs etc..

Rapid progress in electronics and advances in experimental techniques driven by other large detectors
(like liquid argon calorimetry of ATLAS) can influence the decision regarding readout electronics:
better solutions can be had for the same cost or the old solutions can implemented with lower costs.

In this chapter we try to argue that a realistic, proven to work detector can be constructed with well
understood costs, hence we accept all the technological choices adopted by the ICARUS collaboration:
all electronics outside the detector cryostat. Such a detector can be built and it will function, 
as shown by the T600 operation. It does not require any significant R\&D effort, hence it can be used as
a baseline. It is very likely that better and/or cheaper detector can be constructed taking advantage
of the advances in electronics and we discuss several possible improvements in  Chapter 8.

The parameter of the electronics is the noise generated at the input of the first transistor.
For the JFET technology and $1~\mu sec$ shaping time, as used in the ICARUS case,   
this is $ENC~=~(500+2.6C) electrons$, where C is the detector capacitance in $pF$. The figure of merit
if the experiment is $S/N$ for the minimum ionizing track. For the detector described in the preceeding 
sections $S \sim 25,000$ electrons, assuming that the argon is purified to the level corresponding
to the electron lifetime exceeding $10~msec$.

The wire capacitance is $450~pF$ for the collection wires and up to $520~pF$ for the stereo wires.
All the wires are read out on top. A fraction of the stereo wires terminate at the sides of the chamber
and they would require much longer cables thus leading to a substantially higher noise. Three planes
of readout provides an over-constrained measurement, whereas two planes are sufficient to determine the space
point. We propose to to read out only stereo wires terminating at the top of the chamber, thus ensuring
good signal-to-noise ratio. Such a scheme will provide a good stereo measurement over the entire area
of the chamber and will provide redundant measurement over more the $60\%$ of the area. The cable 
length is related to a number of penetrations for feed troughs. We envisage a chimney over $3~m$ to keep
the cable length below $4~m$, hence the cable capacitance below $180-200~pF$. The heat load induced by
a penetration is estimated to be $40~W$ hence the total heat leak of the penetrations will require
an increase of the refrigeration power by abut $5\%$.

The total capacitance of the detector will be of the order of $650-720~pF$ thus giving the expected
noise figure of $2200-2400 electrons$ for collection and stereo wires respectively thus giving the $S/N$ 
figure to be 11.4 and 10.4. While these numbers are very comfortable, they do indicate that the argon
purity levels must not fall much below their design values.

Although  we find the overall design of the ICARUS electronics to be of very good quality and be perfectly
acceptable for a very large detector its implementation with discreet components is very expensive.
It appears relatively straightforward to implement this electronics as an custom-made ASIC to reduce
the price. Another, relatively straightforward improvement could be a change of the technology of the input
transistor from JFET to bipolar or GaAs. It may lead to a reduction of the electronics noise by a factor
of $1.5-2$ and boost the $S/N$ figure accordingly. 

The readout scheme of the putative detector will therefore be the following:  
\begin{itemize}
\item a flat twisted pair cable is connecting the a set of wires to a connector on PC feed-through
\item printed circuit feed through is bringing the signals to the outside of the thermally insulated volume
\item short cable is transporting the cable to the FEE board with ASICs located in a VME crate    
\end{itemize}

The cost of production of 300,000 channels of ASIC-based electronics is estimated to be $\$1~M$, including
the necessary engineering. At such a level it is very likely that the cost of the entire readout system
will be dominated by other costs: cables, connectors, feed-troughs, VME crates.  We estimate that $\$5~M$
should be a safe estimate of the cost of the entire readout.

\section{Data Rates in a Surface Detector}

Signals from all the wires will be digitized with frequency of $2.5~HMz$. Assuming 4 bytes per digitization
it sets the requirement for the data acquisition system to handle data rates of $3~Tbytes/sec$. This is 
well within capabilities of a modern data acquisition systems but it requires an LHC-class DAQ. 

Most of the time buckets, or 'pixels' are empty. An advanced signal processing techniques, already implemented
in the ICARUS electronics (DAEDALUS chip) can perform cluster finding and baseline subtraction thus
reducing the data rate into the DAQ system by a large factor.  The resulting data volume is dominated by the 
signals induced by cosmic muons. The rate of cosmic muons crossing entering the detector is about $200~kHz$.
Most of them range out in the upper part of the detector. Their angular distribution leads to a relatively
few wires being hit. This is, in fact, a significant factor to be taken into account in the specification
of the front end electronics, as it leads to a very large dynamic range of the input signals. As a safe
upper limit we assume that the average number of wires hit by a cosmic muon is less than 2000. The expected
data rates is therefore less than $1.6~Gbyte/sec$.
 Muon induced electromagnetic cascades can produce higher occupancy but they are infrequent enough 
not to contribute to the average data rates.

In an accelerator experiment with external beam timing (providing  a time reference, $T_{0}$)
a beam duty  factor reduces the data volume by a factor of $10^3$. 

We envisage the detector to be a tool for a wider range of physics topics than the appearance experiment, hence
we a more powerful data acquisition system is appropriate. The eventual cost is likely to be dominated by
the cost of data storage and distribution system. Major savings can be realized with intelligent data 
compression techniques or with removal all of the recognized cosmic muons from the data sets.

We estimate that the cost of the data acquisition system and the necessary data storage systems will
be of the order of $\$10~M$.

\section{Total Detector Cost}

The detailed cost estimate does require a more detailed engineering studies. The detector described
here is of un-precedented scale on one hand, but it is relatively simple on the other hand.
A significant fraction of the total cost are the industry-provided items (tank, argon, refrigeration
systems, purification systems) whose costs are relatively well known as the scale of our application
is quite typical in  the industrial systems.

\begin{table}
\begin{center}
\begin{tabular}{|l|r|}
\hline
Detector subsystem & estimated cost in \$M \\ \hline
Cryogenic tank                  & 11.0   \\ \hline
Liquid Argon(delivered)         & 30.0   \\ \hline 
Refrigeration and Purification  & 12.0   \\ \hline
Structural Supports             & 37.7 \\ \hline
High Voltage and Field Shaping  & 5.7 \\ \hline
Inner Detector                  & 5.0     \\ \hline
Electronics                     & 5.0     \\ \hline
Data Acquisition and Storage    & 5.0     \\ \hline
TOTAL                           & 111.4    \\ \hline

\end{tabular}
\end{center}
\caption{Estimated cost of the 50 kton Liquid Argon Off-axis Detector.}
\label{cost}
\end{table}

This cost may be contrasted with the construction cost of $1.2~kton$ ICARUS module, which is 
of the order of $\$25~M$. The difference is huge and must be understood and explained. It is 
largely due to the difference related to the space and safety requirements of the Gran Sasso
experiment and to the advances of the electronics industry as well as to the  advantages 
of the large detectors in terms of the volume to surface ratio. As an example, a total number
of wires for 40 times bigger Off-axis detector is only five times larger that that in the T1200. 
It is instructive to point out
that the cost of the liquid argon is the dominant cost of the large detector is a minor cost
element of the T1200.

Here the main differences between the Off-axis detector and the T1200:
\begin{itemize}  
\item large industrial cryogenic tank vs thin-walled, custom cryostat with the active 
thermal shield. The cryostat and associated cryogenics costs cover about $40\%$ of the
T1200 module.
\item discreet components electronics (the second largest cost element) is a very expensive
solution. Advances in the electronics area make custom ASIC much more economical solution.
The scale of the necessary readout electronics is very similar to the electronics of the upgraded
silicon detectors of D0 and CDF, hence the cost estimate is probably very realistic.
\item large volume of the detector allows for simple mechanical solutions for the inner detector
system, thus reducing the construction costs considerably.
\end{itemize}  

On the other hand it appears surprising that the cost of the high-performance imaging detector
appears to be comparable or even lower that that of an inferior sampling calorimeter detector.
This can be understood as a being do to the following factors:
\begin{itemize}
\item a detector hall for a very large detector has a significant cost associated with it $\$25-30~M$.
This is substantially higher that the cost of the storage tank.
\item the cost of the liquid argon is comparable to a OSB absorber and within $50\%$ of the cost of 
of the particle board.
\item a number of detector channels is comparable or actually smaller in case of the liquid argon
owing to a long drift distances. These detector channels, wires, are relatively simple yet they
provide far more detailed information than  the traditional detector.
\end{itemize}  
     
\section{Detector Site}

To first order, the optimal site location for neutrino oscillation detector
is that which maximizes $\sin^2{1.27\frac{\Delta m^2L}{E}}$. 
For the present best value of the atmospheric $\Delta m^2$, 
the oscillatory term is maximal for an $L/E$ of 450 to 500 km/GeV. 
Matter effects favor a longer baseline ($L$) while
flux considerations (number of events) favor a shorter baseline. The NuMI
off-axis beam at an off-axis angle of $\sim$13.7 mrad exhibits the best
combination of maximum flux and sharp peaking in energy. This corresponds to
a mean beam energy of around 2 GeV giving an ideal baseline of 850 to 900 km.

Further desirable characteristics for the experiment site are: 
\begin{itemize}
\item access by rail to minimize transportation costs,
\item proximity of a highway that is maintained year-round,
\item stable local geology, 
\item location not too close to a populated area,
\item availability of a wide lateral area (i.e., spanning a range in off-axis angle)
so that the choice of exact location can be made to optimize the physics potential.  
\end{itemize}

A location which satisfies all these criteria is 
located near the western boundary of the Canadian province of Ontario. The
nearest town, Fort Frances, is about 60 km west and borders the American
city of International Falls. The small village of Mine Centre is about 
15 km east of the site. The central NuMI beam axis is about 8.7 km above
the surface at this point and the site for the 13.7 mrad 
off-axis beam is then about 8.6 km east of the central beamline. 
Just as an illustration of the siting flexibility associated with
this location, we notice that there is also a suitable site at the 12 
mrad off-axis.  Sites to the west of the central beamline would be located
in a First Nations reserve and so were not considered further. 

All sites 
to the east of central beamline are on crown (public) land and so land
use should not be an issue. It is possible to locate the detector quite
close to the railroad. Depending on how close it is feasible to get, 
rail spur costs can be as low as several hundred thousand dollars. All sites
are a few km north of Trans-Canada Highway 11 which is a main highway
maintained year-round.

The major port city of Thunder Bay is around 360 km east allowing for the
possibility of transporting the liquid argon by ship most 
of the way and then transferring it to a train for final transport to
the site. The basic underlying geology of the
region is granite, hence   the area is very 
stable geologically.

\chapter{Small Liquid Argon Detectors}

Experience gained with successful construction and operation of the T600
module of ICARUS allows for a relatively straightforward design of smaller
detectors suitable for Fermilab's neutrino beams. They can be placed in the near MINOS hall
or in a dedicated new experimental hall in the MiniBOONE beam. As an example we describe
a possible design of a 40 ton detector, T40\cite{Franco_T40}, but with a relatively
straightforward extrapolation such a design can be scaled to detectors in the range
$20-100~ton$.

\section{T40 Detector} 
\begin{figure}[t!]
\begin{center}
\includegraphics[scale=0.4]{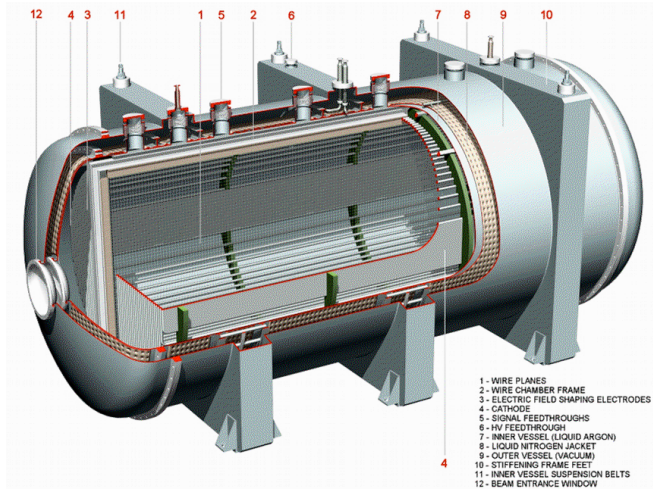}
\vspace{0.5cm}
\caption[]{ T40 Detector: cutaway view. }

\label{T40}
\end{center}
\end{figure}

The detector is constructed as a cylinder filled with liquid argon (see Fig.\ref{T40}). 
A vertical wire chamber (item 1) is 
aligned along the cylinder axis and splits the volume into two drift regions 
(left and right). 

A horizontal and uniform electric field is generated 
in each region by a cathode (4) and a system of field shaping electrodes(3).
 
The chamber is built of two   sets of wire planes. An additional central  grounded
 wire plane screens the left and the right sets one from the other.

Signals from wires are brought to the front-end electronics, outside the 
cryostat, through twisted pair cables and low voltage feedthrough flanges (5).
 Two high voltage feedthroughs (6) are used to bias the cathodes and the field
 shaping electrodes.

The vacuum insulated and liquid nitrogen cooled cryostat is built of an 
inner vessel and an outer vessel, with vacuum in between.
The inner vessel is surrounded by a dimpled surface jacket welded to it and 
refillable with LN2. Three square frames are part of the outer vessel. They 
are used as bases for it and as supporting frames for the inner vessel. 
Wrapping the inner vessel and the LN2 jacket by super-insulation layers and 
suspending them to the outer vessel frame by three steel belts (11) minimize 
the heat input.

\begin{table}
\begin{center}
\begin{tabular}{|l|r|}
\hline

Active liquid argon sizes &	$\Phi$ = 2.4 m, L=6.0 m   \\ \hline
Active liquid argon mass    &	38.5	Ton   \\ \hline
Number of drift regions    &	2   \\ \hline	
Max drift lengths   &	2 × 1.2	m  \\ \hline 
Maximum required high voltage   & 	60	kV \\ \hline  
Number of cathode planes    &	2   \\ \hline 	
Number of wire chambers	 &   1  	\\ \hline
Number of readout wire planes  & 	4   \\ \hline	
Orientation of readout wires  & 	0°, 90	\\ \hline  
Number of readout wires	  &  5632   	\\ \hline
\hline
Heat Input	 &  \\ 
	a) Radiation (with wr = 1 watt/m2)   &	72	W  \\ 
	b) Conduction (cables + mech. supports)   &	240	W  \\ 
	Total   &	312	W   \\ \hline
	Equivalent liquid nitrogen consumption   &	0.2	$m^3/d$   \\ \hline 
\hline
 Dipole magnet (warm, B = 0.6 T)  &	\\
	Copper weight   &	60.2	Ton     \\ 
	Power @ 35 kA   &	2.5	MW    \\ 
	Iron yoke    &	900	Ton  \\ \hline   

\end{tabular}
\end{center}
\caption{T40 - Parameters}
\label{near_det}
\end{table}

The main parameters of the T40 detector are reported in Table\ref{near_det}. 
The low 
nitrogen consumption allows operating the detector with a reduced size, 
refillable nitrogen storage dewar.

A relatively small size of the described detector with respect to the T600
module of ICARUS allows for significant simplifications of the construction and 
cryogenics, vacuum insulation being the prime example.

Construction cost of such a detector can be reliably estimated from the
experience of the ICARUS module construction\cite{Franco_cost}.

\begin{table}
\begin{center}
\begin{tabular}{|l|l|r|}
\hline

Liquid argon & $60 m^3$ & \$30K \\ \hline
Cryostat  & Inner vessel & \\
  &	LN2 jacket & \\ 
  &	Outer vessel & \\ 
  &	Beam entrance windows & \\ 
  &	Chimneys for signals, HV, & \\ 
  &	IN/OUT Ar, IN/OUT LN2,  & \\ 
  &     inner vessel suspension  & \$500K \\ \hline 
Inner detector  & Wire chamber (frame, wires) & \\
&  Field shaping electrodes & \$70K \\ \hline 
Electronics (5632 ch.) and DAQ & Signal cables & \\
  &    feedthroughs & \\
  & Analog and Digital  crates  & \\
  & Calibration pulser & \\
  & Wire bias HV power supplies & \\
  & Data acquisition  computer & \$350K \\ \hline 
Vacuum \& cryogenics &  Vacuum pumps and gauges & \\
  & LN2 and LAr storage dewars & \\
  & LAr purification system and purity monitor & \\
  & Transfer lines, valves & \\
  & Level and temperature monitors & \$75K  \\ \hline 
High voltage system & Power supply, feedthrough, monitor & \$25K \\ \hline 
Other systems & External trigger counters and electronics & \\
  & UPS and O2 monitors & \$150K \\ \hline 
Total & & \$1200K \\ \hline

\end{tabular}
\end{center}
\caption{ T40 - cost estimate}
\label{near_cost}
\end{table}

\begin{figure}[t!]
\begin{center}
\includegraphics[scale=0.4]{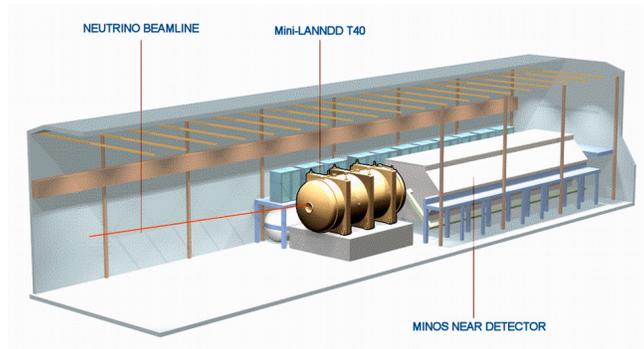}
\vspace{0.5cm}
\caption[]{T40 detector located in the near MINOS hall. The near MINOS
detector serves as a muon catcher. The detector is raised to the nominal
height of the incoming neutrino beam. }

\label{T40_near_hall}
\end{center}
\end{figure}

These cost estimates, Table \ref{near_cost} are valid for a construction in a surface hall. They
are probably accurately estimate of a construction in a possible new hall
in the MiniBOONE beam. A construction of a possible T40 class detector in 
a near MINOS hall (Fig.\ref{T40_near_hall} would be complicated by the size 
of the NuMI access shaft. In this case the cryostat needs to be constructed 
in place, nearly doubling the estimated cost of the cryostat. 

\section{Near Liquid Argon Detector in an Underground Hall}
 
An operation of a liquid argon detector in an underground hall requires
careful analysis of the installation and operation and the safety aspects.

To put the upper limit on the associated costs we have considered the case
of the operation in the near MINOS hall, some $100~m$ underground\cite{schmitt}.
Installation and operation in a dedicated hall in the MiniBOONE beam would be
considerably easier and cheaper, but such an experimental hall would have to be constructed.
 
Cooling would be provided by liquid nitrogen piped through vacuum insulated lines
from the surface and the gaseous nitrogen returned to the surface. There would be
no liquid argon storage on the surface: the tank would be filled directly from 
the supply trucks. If necessary  the argon  would pumped  to the 
surface and vented. A detailed study would be required to evaluate and mitigate
the oxygen deficiency hazard, but an it is not expected to be a major issue.

Underground operation will lead to significant  costs of the additional 
equipment, installation and testing  and analysis of the safety requirements.

These costs are estimated to be \$250 K for materials and services, 70 man-weeks
of engineering effort and 178 man-weeks of technician effort. 

\section{Magnetic Field?}

The possibility of operating liquid argon TPC's in magnetic field is, by far, 
the most interesting implementation of this kind of detector. This feature 
allows easy muon sign discrimination and momentum evaluation. At beam momenta 
lower than 5 GeV/c, with a magnetic field intensity of 0.6-1.0 T, calculations
 indicate that the sign discrimination for e± that initiated an 
electromagnetic shower becomes possible too. The main parameters for a 
possible configuration with a dipole magnet are indicated at the end of 
Table\ref{near_det}.

Such a configuration would enhance the physics potential of T40 detector
for the studies of low energy neutrino interaction and it would be of major
interest for the future experiment at  neutrino factories. Given the cost and
associated complications one would probably consider addition of a magnet as 
a second phase of the possible experiment.
   
\chapter{Double Beta Decay Experiment with a Liquid Argon Imaging Detector}

 \section{Double beta decays and neutrinos}
 
 Interest of neutrino-less double beta decays ( 0$\nu \beta \beta $) has recently been sharpened since they may have discovery potential for the Majorana neutrino masses and other fundamental properties of neutrinos \cite{eji02}. 
 
 Fundamental neutrino properties to be studied by 0$\nu \beta \beta $) are the Majorana nature of neutrinos, the type of mass spectrum, the absolute neutrino mass scale and possibly the CP violation. Actually, 0$\nu \beta \beta $ experiments is the only practical method for studying all these important properties of neutrinos in the foreseeable future. 

  The data from Super-K, SNO and KamLAND, together with their theoretical analyses, clearly imply that next-generation experiments with the mass
sensitivities of the atmospheric $\nu $-mass scale (50 meV) should discover non-zero effective Majorana electron neutrino mass if the massive neutrinos are Majorana particles and the neutrino mass spectrum is of the quasi-degenerate type or with inverted hierarchy. In fact, many theories of the fundamental particle interactions predict that the massive neutrinos are Majorana in nature.  Experiments with even higher sensitivities of the solar $\nu $-mass scale(a few meV) are of potential interest for further studies in case of normal hierarchy and possibly of the CP violation.

Since 0$\nu \beta \beta $ events are extremely rare low-energy processes, the experiments require large amount( $\sim $ ton) of $\beta \beta $ isotopes, large-scale low-background detectors and very stringent selection of the signal from background events. The 0$\nu \beta \beta $ rate depends on the nuclear matrix element as well. Accordingly it is indispensable for identifying the $0\nu \beta \beta $ event to perform at least  two or three independent experiments utilizing different isotopes and methods. Here the different methods should include the calorimetric method with active source detectors and the spectroscopic $\beta $-tracking method with external sources.

\section{MOON}

The nucleus $^{100}$Mo has large responses for both the $\beta \beta $ decay
and low-energy solar and supernova $\nu $'s \cite{eji01}. Thus Majorana neutrino masses and the charged-current interaction of solar and supernova $\nu $'s can be studied effectively using $^{100}$Mo. 

The MOON (Molybdenum Observatory Of Neutrinos) project \cite{eji00} is a
hybrid $\beta\beta$ and solar $\nu$ experiment with planned sensitivity to
Majorana mass of the order of $<m_{\nu}>\sim$0.03 eV as well as capability for
spectroscopy of {\em pp} and $^7$Be solar $\nu $'s. 

The two $\beta$ rays from $^{100}$Mo are measured  in prompt coincidence  for the 0$\nu\beta\beta$ studies.
 The large $Q$ value of $Q_{\beta\beta}$=3.034 MeV gives a large
   phase-space factor $G^{0\nu}$  to enhance the 0$\nu\beta\beta$ rate, and the
energy and angular correlations for the two $\beta$-rays are used to identify
the $\nu$-mass term for the 0$\nu\beta\beta$.

 Low energy solar-$\nu$  captures are measured in delayed coincidence with the
subsequent $\beta$ decay of the 16-second daughter $^{100}$Tc. The 
low threshold energy of 0.168 MeV for the solar-$\nu$ and the large capture rates allow real-time studies of pp and $^7$Be $\nu $'s. With a low threshold energy and large GT and spin-dipole strengths, $^{100}$Mo is also sensitive to electron $\nu $'s from supernovae.

   The measurement of two $\beta$-rays (charged particles) enables one
   to localize in space and in time the decay-vertex points for both the
0$\nu\beta\beta$ and solar-$\nu$ studies. The localization is crucial  for
selecting $0\nu\beta\beta$ and solar-$\nu$ signals and for reducing  correlated
and accidental backgrounds.

  MOON is required to have a large amount ($\sim $ one ton ) of $^{100}$Mo
isotope (whether natural or enriched) and good energy and position resolutions.
Research and development of possible detector options are in progress. They
include supermodules of plastic fiber scintillators,  liquid scintillators, 
cryogenic bolometers and others. The Mo source and scintillator are to be purified to the level of1 ppt \cite{rob98}. 

\section{Double Beta Decay Experiment with Liquid Ar}
Liquid Ar detectors have several merits for studying 0$\nu \beta \beta $. 
We briefly discuss the energy resolution, the position resolution, and the background rate, which are crucial for high sensitive studies of double beta decays.

\begin{figure}[t!]
\begin{center}
\includegraphics[scale=0.3]{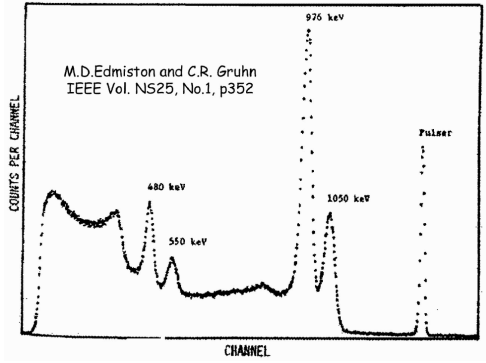}      
\vspace{0.5cm}
\caption[]{The pulse height spectrum of $^{137}Cs$ measured with liquid argon ionization chamber.
The energy resolution, limited by electronics, is $\Delta E/E = 0.015$ at $E= 976~keV$  }          

\label{lar_res}
\end{center}
\end{figure}
The energy resolution of the liquid Ar detector itself is  very good as shown in 
Fig.\ref{lar_res} \cite{energy_resolution}. A significant contribution to the energy resolution is due to 
the energy loss in the Mo foil. Noting that the two beta rays from the 0$\nu \beta \beta$ event are emitted 
opposite with each other, we measure such 2 $\beta $ rays above 0.5 MeV. Then the energy spread of the  
sum of the $\beta \beta $ rays is given as FWHM  = $\delta E (\sqrt 2 - 1)$, where $\delta E$ is the 
foil thickness. Using the Mo foil with 30 mg / cm$^2$, one gets FWHM = 20 keV, which is only 0.7 $\%$. 
Thus it is possible to carry out a high resolution (FWHM $\le 1 \%$ experiment even with the Mo foil.

The major BG of the liquid Ar detector comes from the impurities of the radioactive $^{42}$Ar. It decays with $T_{1/2}$ = 33 y and $Q$ = 0.6 MeV to $^{42}$K, which decays successively to $^{42}$Ca with T$_{1/2}$ = 12.4 hr and Q = 3.52 MeV The $\beta $ decay with 81$\%$ branch to the ground state can partly invade into the $0\nu \beta \beta $window.

The BG rate per year per 1 ton $^{100}$Mo is evaluated as 
\begin{equation}
Y(^{42}K) = a M \epsilon ~2.6 10^{26}
\end{equation}
where $a$ is the $^{42}$Ar impurity ratio , $\epsilon$ is the efficiency and $M$ is the mass ratio of Ar to Mo. $M$ can be $M \sim 30$ to stop most of $\beta $ rays below 2 MeV. The efficiency is estimated to be $\epsilon \sim 2 10^{-5}$. It is noted here that most of the $\beta $ rays from $^{42}$K with the half-life of 12.4 hr can be rejected by looking at the preceding $\beta $ decay from $^{42}$Ar by using adequate segmentation. Thus the GBG rate per year ton is  $Y(^{42}K) = a ~1.5~ 10^{23}$. 

 It should be  noted that most RI impurities( natural
and cosmogenic) do not come into the 0$\nu \beta \beta $ window because of
the quite high $Q_{\beta \beta }$.  The $^{214}$Bi $\beta $ can be rejected
by measuring the post $\alpha$ decay after 100-200 micro sec.

The $^{42}$Ar impurity is reported to be $a \le 10^{-21}$. Then one gets $Y(^{42}K) \le 1.5 ~10^{2}$. This corresponds to the neutrino mass sensitivity of the order of 50 meV for 5 year measurement with 1 to $^{100}$Mo. Thus we need R$\&$D to reduce further the impurity of $^{42}$Ar to the level of $\le 10^{-22}$ in order to achieve the mass sensitivity of the order of 30 meV.

The $2\nu \beta \beta $ contribution to the $0\nu \beta \beta $ window can be negligible in the present case of the energy resolution of FWHM = 1 $\sim  2 \%$.

The position resolution is required to reduce the 2$\nu \beta \beta $ BG at the solar neutrino window. The inverse $\beta $ ray from the $^{7}$Be solar neutrino is measured in the delayed coincidence with the successive $\beta $ decay of $^{100}$Tc. Using the time window of 30 sec, The accidental rate at the $^7$Be window is 
\begin{equation}
Y(\beta \beta ) = \epsilon ' ~2.6~ 10^7~ K^{-1}
\end{equation}
 where $\epsilon '$ is the efficiency and $K^{-1}$ is the position resolution in unit of ton. Assuming a modest efficiency of $\epsilon ' \sim 5 \%$, the position resolution of around $K^{-1} = 10^{-7}$ is required. This corresponds to 0.1 g of $^{100}$Mo, i.e. 2 cm $\times $ 2 cm for the foil with 30 mg/cm$^2$. This is well within the capabilities of the liquid 
argon detector. 

The  liquid argon TPC detects two
individual $\beta$- tracks emitted opposite to each other to identify the
0$\nu \beta \beta $. This is unique among other high sensitive
 calorimetric experiments.

\section{Liquid Argon Detector}

The detector takes advantage of the Mo foils using them as cathode planes. The drift distance required to 
contain fully the electron tracks is $2~cm$, hence the argon purity requirements are relaxed by about two
orders of magnitude compared with the ICARUS module.
 
The fundamental
detector cell consists of a cathode plane and two wire chambers (two wire planes, X and Y, each) on either side of the
cathode for a total thickness of $5.5~cm$. To ensure required position resolution the wire pitch is $1~ cm$.
As every cell provides a complete measurement of the $0\nu \beta \beta $ decay the detector can be constructed 
in a modular fashion, if so desired.

A detector for $1~kton$ of enriched molybdenum appears relatively straightforward (especially in comparison
with the enrichment itself. It would be somewhat smaller than ICARUS test module and it would contain about
$400 ton$ of liquid argon. The main parameters of a possible monolithic detector are summarized in Table\ref{db_det}.

\begin{table}[t]
\begin{centering}
\begin{tabular}{||l|r||}\hline\hline
Cryostat  cross section (inner) & $4.5\times 4.5~m^2$  \\ \hline
Cryostat length   & $14~m$   \\ \hline
Mass of liquid argon & $397 t$ \\ \hline
Cathode plane ($^{100}Mo$) thickness  & $25~\mu$ \\ \hline
Cathode plane size & $4 \times 4~m^2$ \\ \hline
Number of cathode planes & 250 \\ \hline
Wire chamber dimensions & $4 \times 4 ~m^2$ \\ \hline
Number of wires planes/chamber & 2(X and Y) \\ \hline
Wire pitch & $1 cm$ \\ \hline
Total number of chambers & 500 \\ \hline
Total number of wires & 400,000 \\ \hline
Total number of electronics channels & 400,000

  \\ \hline \hline
\end{tabular}
\caption{Parameters of $1~kton$ $^{100}Mo$ double beta decay detector}\label{db_det}
\end{centering}
\end{table}

This detector is very similar, conceptually,  to the one operated in Gran Sasso laboratory \cite{barabash}.
Such a detector can be constructed using the existing technology, the prime
challenge being the molybdenum enrichment. The most critical detector 
performance parameter is the energy resolution which is, turn, limited by
electronics noise, which is related to the detector capacitance. The wire
capacitance is minimized by relatively large wire spacing and little 
improvement can be expected here.

Possible development of a new amplifier using other than JFET technology
is one possible avenue.  The biggest improvement can be attained by development
of a cold version of the readout electronics which is mounted directly on the 
chambers inside the cryostat. Such a solution would eliminate the cables,
which are the dominant contribution to the source capacitance. Low operating
temperature would reduce further the electronics noise.

\chapter{Proposed R\&D Program}

In the preceeding chapters we have argued that a liquid argon technology
with the very fine spatial granularity, good energy resolution, good
particle detection and identification  and imaging capabilities 
opens a qualitatively new phase in neutrino experiments.  

This is a mature technology. Detectors can be constructed based on the
knowledge we have in hand and they can be surprisingly inexpensive. 
It does sound too good to be true. The primary goal of the proposed
R\&D program is to establish a credibility of this proposition.
Once this credibility of the technical claims is established we envisage
a number of possible improvements to the existing technology, thus
making the detectors cheaper for a given performance, better for a given
cost or more reliable.

A successful completion if the  R\&D activities described below is necessary to
make credible and defendable proposals of the future experiments. Total required
resources are:

\begin{itemize}
\item 12 months of engineering support
\item $200~K$ for materials and services
\item space and technical support for the construction of the small detector
prototype
\end{itemize}

\section{Establish Credibility of the  Experimental Proposals}

In this letter we have made two assertions:
\begin{itemize}
\item high-performance liquid argon ``Near Detector'' of $40~kton$ class can
be constructed by simplifying and scaling down the ICARUS T600
module. Such a detector can be constructed in a relatively short 
time scale and it would offer a very attractive neutrino physics
program

\item a very large surface detector can be constructed using 
existing liquid argon technology combined with the industrial
techniques for cryogenic liquid storage, refrigeration and
purification
\end{itemize}

Given the potential payoff in terms of the physics potential we think
it is very important to demonstrate credibility of these claims. This
is the primary goal of the proposed R\&D program.

\subsection{'T40'  Detector}

The detector is a simplified and scaled down version of the ICARUS
module. The main differences are:
\begin{itemize}
\item small, simplified vacuum insulated cryostat
\item cylindrical rather than rectangular drift volume (this should
have little bearing on the performance and construction though). 
\end{itemize}

All of the technical aspects of such a detector are already demonstrated
by the ICARUS collaboration. Scarcity of the first technical expertise 
in the US is the chief obstacle in making a detector proposal. We propose
to address this by a construction, in collaboration with the ICARUS
colleagues, of a small TPC module as a means of the 'technology transfer'.

A successful operation of such a chamber, expected to be in the $100~l$
of liquid argon, $80~cm$ drift distance range, will provide an important
proof of the relative maturity of the technology itself. Such a chamber
is under construction using bits and pieces borrowed from ICARUS
and from various Fermilab shops. To maximize the chance of a success this
project requires some $\$100 K$ to procure some cryogenic elements, 
purification filters etc.. With the collaboration of the ICARUS colleagues
we hope to bring this prototype into operation no later than the spring
of 2005.

A proposal for an experiment using small LaR TPC requires a detailed safety
analysis related to possible hazards of operation in confined, perhaps
underground locations and the related cost implications. For the case of
an experiment located in the near MINOS hall additional costs of an
undergrond construction must be analyzed. An initial analysis shows that
all the problems can be addressed with no serious cost implications,
but we expect that some 3 months of engineering work is necessary to
have full confidence in the conclusions.

\subsection{50 kton Detector}

This is an extrapolation of the ICARUS module based of the established facts:
\begin{itemize} 
\item liquid argon can be purified to levels enabling drift distances up to $3~m$
\item front-end electronics can be build to ensure good signal-to-noise ratio
for detector capacitances up to $400~pF$
\end{itemize}

The primary challenge in scaling up to much larger detector volume is to
design a cost-effective detector while maintaining the detector performance.
The latter reduces mostly to ensuring sufficient liquid circulation
capacity through a set of commercial filters. 

Cost effective solution for the large detector design is accomplished by
replacing modular custom-made cryostats of ICARUS with an industrial
grade liquified gas storage tank. This replacement  provides the most of the cost savings, 
but it leads to a natural question if such a tank will permit the  required
purity level of liquid argon to be achieved. It is a very relevant question
as the large tank solution requires that evacuation as an initial stage
of cleaning  must be replaced by several purges with nitrogen and argon.

The question of the ultimate argon can only be answered 
in situ with the ultimate detector, but a high level of confidence can be
achieved by detailed engineering studies of the material purities and the
purification process. To this end we plan to use the small test chamber mentioned 
above to investigate the relevance  of various cleaning procedures as well
as the influence of materials likely to be used in the final detector on the
argon purity level. It is expected that the studies of the purification 
process will involve industry producing purifying filters.
One of the suggested tests may involve purification of 
the liquid argon in one of the commercial nitrogen storage tanks available
at the Fermilab site. We expect that the studies of the purification process 
and the achievable purity levels will require 3 months of engineering and $\$100~K$.

Apparent simplicity of the ultimate detector suggests that the cost estimates may
be fairly realistic. The devil is in details, though, and it may be that significant
costs may be hidden in the details ignored so far. We estimate that 6 months of the
detailed engineering effort is necessary to produce very reliable material and 
construction cost estimate.  


Very significant cost savings, with respect to the ICARUS module, can be realized 
by implementation of the front end electronics in a customized ASIC. We consider
this step to be well understood and not requiring a proof of feasibility.
On the other hand, various further improvements in the electronics are desirable
for the real experiment and they will be described later.
  
\subsection{Double Beta Decay Detector}

Feasibility of the liquid argon-based double beta decay experiments depends 
on the background levels attainable with sheet geometry on one hand and induced
by $Ar^{42}$ on the other hand. These background calculations will be carried out
by Osaka and University of Washington groups. 

Achievable spatial and energy resolution will be limited by the electronics
noise of the detector. Optimization of the wire  readout system, for example multiple
induction planes to reduce the electronics noise, will be carried out with the 
prototype mentioned earlier.

A significant improvement can be achieved by further improvements of the 
low noise, high gain preamplifiers.

\section{Future/Further R\&D Efforts}
Having all the results of the ICARUS efforts and profiting from a close collaboration
with this group we estimate that this phase of the R\&D can be carried out within one
calendar year and we expect it will lead to proposals of specific physics experiments.

Being confident that these experiments will be judged important enough to become a part
of the scientific program of the Laboratory we envisage further program of R\&D
aimed at the further improvements of the detector technology and data analysis techniques.

These studies will include:
\begin{itemize}
\item studies of other technologies (bipolar or GaAs) for the front end electronics to 
reduce the input  noise and to improve signal-to-noise ratio.
\item studies of a possible solution for cold electronics (i.e. electronics inside
the cryostat) to reduce further the electronics noise by elimination of the cable
capacitance and to reduce or eliminate problems related to cables (out-gassing, feedthroughs, etc).
\item optimization or simplification of the front end electronics using modern industrial
solutions (for example modern FADC for every channel instead of the multiplexing, FPGA arrays
for the cluster finding, etc..)
\item investigation of data analysis and event reconstruction techniques for four  dimensional
(including  floating T0) data
\item investigation of data compression, data organization and data storage technologies 
for the massive amount of data expected from the free-running surface detector
\end{itemize}



\newpage
\appendix
\renewcommand{\thepage}{\Alph{chapter}-\arabic{page}}

\end{document}